# Introduction to SQUID Sensors for Electronics Engineers

Yongliang Wang [a,b*]

[a] *Shanghai Institute of Microsystem and Information Technology (SIMIT), Chinese Academy of Sciences (CAS), Shanghai 200050, China*
[b] *CAS, Center for Excellence in Superconducting Electronics (CENSE), Shanghai 200050, China*

[*]Corresponding author. Tel.: +86 02162511070; Fax: +86 02162127493.

E-mail address: wangyl@mail.sim.ac.cn
ORCID: Yong-Liang Wang (0000-0001-7263-9493)

**Abstract**

Flux transformers and superconducting quantum interference devices (SQUIDs) are two key superconducting devices used in the cutting-edge magnetic-field sensor systems, such as magnetocardiography (MCG) and magnetoencephalography (MEG). They are superconductor integrated circuits enabled by two superconductor elements, superconductor wire and Josephson junction, based on Meissner's effect and Josephson effect. The principles of SQUID sensors are unfamiliar for electronics engineers who are trained with semiconductor electronics, due to the superconductor physics. We present an introduction to SQUID sensors, ranged from superconductor elements to the superconductor-semiconductor hybrid system. It is demonstrated that, SQUIDs inside are the resistor-inductor-capacitor (RLC) network driven by Josephson currents, and share the common network equation with normal RLC circuits; SQUID outside are magnetic-field-effect transistors (MFETs), and are dual to the semiconductor FETs in signal conversions; SQUID magnetic-field sensors are linear current amplifiers, as semiconductor FET electric-field sensors are linear voltage amplifier.

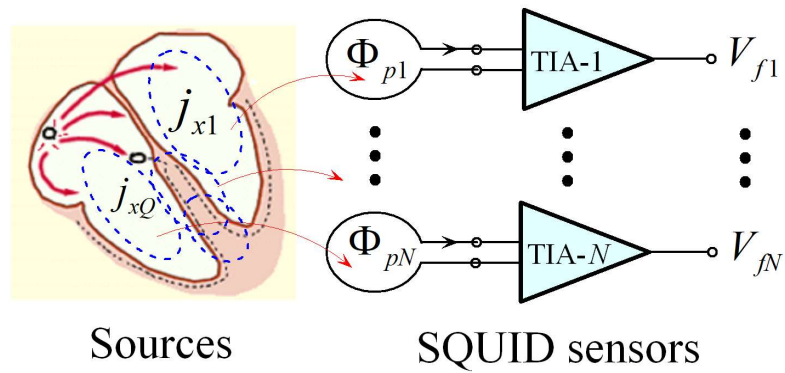

## Elements and Devices

(a) Superconductor  (b) Josephson junction  (c) Flux-transformer

(d) rf-SQUID  (e) dc-SQUID  (f) bi-SQUID  (g) dro-SQUID

(h) SQIF  (i) SQUID-array

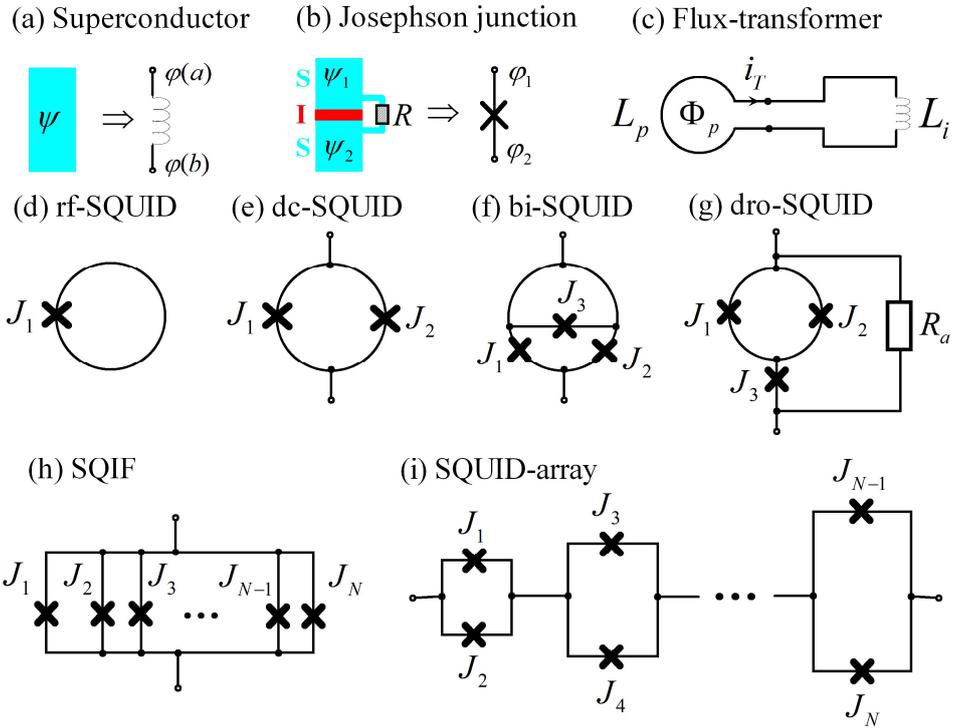

## Levels of abstraction

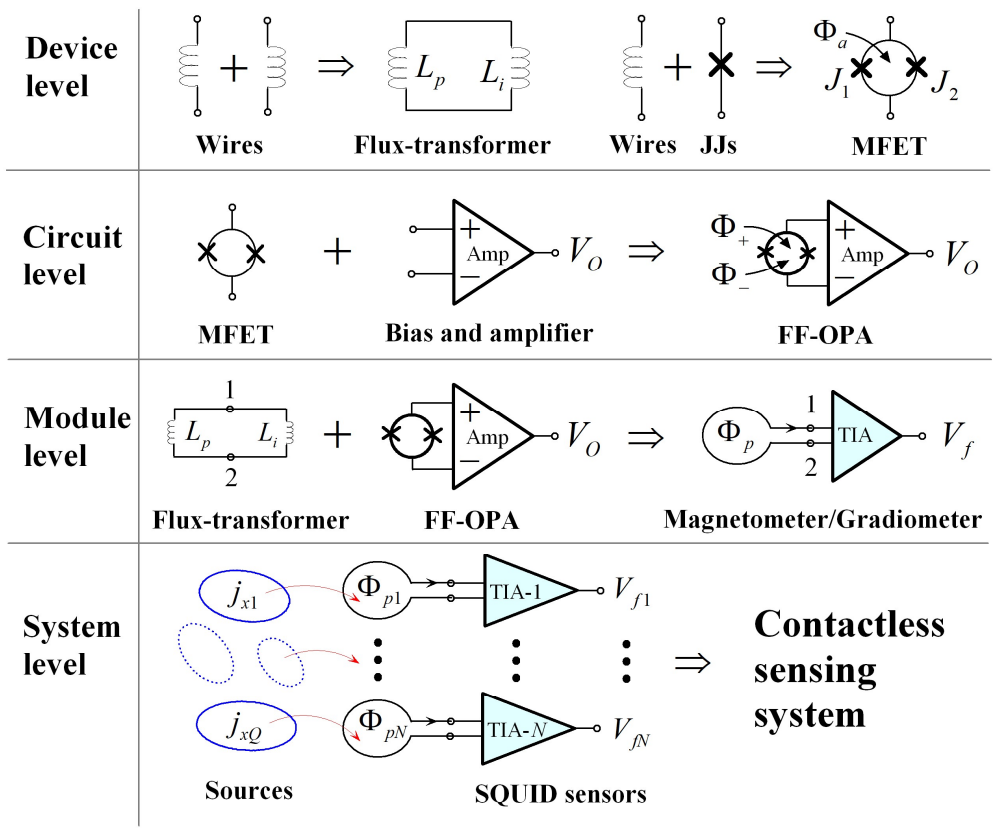

## Overview of SQUID electronics

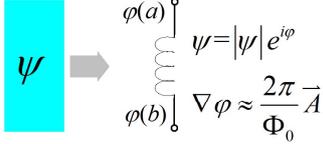

## Duality between MOSFET and SQUID electronics

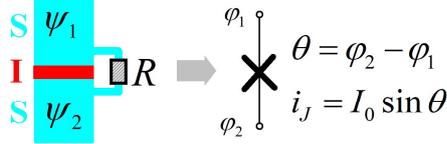

## Abbreviations

**SQUID**: Superconducting Quantum Interference Device
**SFQ**: Single-Flux Quantum
**SQIF**: Superconducting Quantum Interference Filter
**MCG**: Magnetocardiogram
**MEG**: Magnetoencephalogram
**FTMG**: Full Tensor Magnetic Gradiometer
**JJ**: Josephson Junction
**RLC**: Resistor-Inductor-Capacitor
**RCSJ**: Resistively-Capacitively-Shunted Junction
**FQL**: Fluxoid Quantization Law
**KCL**: Kirchhoff's Current Law
**KVL**: Kirchhoff's Voltage Law
**VCO**: Voltage-Controlled Oscillator
**MFG**: Magnetic-Flux Generator
**MFF**: Magnetic-Flux Flow
**MFET**: Magnetic Field Effect Transistor
**CI-MFET**: Current-Input MFET
**MOSFET**: Metal-Oxide Semiconductor Field-Effect Transistor
**AFE**: Analog-Front-End
**GBP**: Gain-Bandwidth-Product
**FLL**: Flux-Locked Loop
**OPA**: Operational Amplifier
**VF-OPA**: Voltage Feedback Operational Amplifier
**CF-OPA**: Current Feedback Operational Amplifier
**FF-OPA**: Flux Feedback Operational Amplifier
**INA** : Instrumentation Amplifier
**TIA** : Trans-impedance Amplifier
**CBVA** : Current-Bias-Voltage-Amplifier
**VBCA**: Voltage-Bias-Current-Amplifier

## Variables

$\psi$: macroscopic waver function;
$\varphi$: macroscopic quantum phase;
$A$: vector potential of magnetic field;
$\Phi_0$: flux quantum; $\theta$: phase difference;
$u$: voltage difference; $I_0$: critical current;
$V_s$: average voltage of dc-SQUID
$\omega_{rf}$: resonance frequency;
$i_J$: Josephson current;
$i_{cir}$: SQUID loop current;
$\sigma_i$: factor of incidence matrix;
$\Phi_m$: total flux in loop;

$\Phi_a$: applied flux; $i_n$: noise current

$L_m$: loop inductance;

$L_p$: inductance of pick-up coil;

$L_i$: inductance of input coil;

$n_T$: number of flux quanta trapped in flux-transformer;

$\Phi_p$: flux picked by pick-up coil;

$I_b$: external current source;

$i_b$: bias current;

$i_1$: current through $J_1$;

$i_2$: current through $J_2$;

$G_{cmm}$: equivalent conductance of two Josephson currents;

$V_{s0}/\Phi_0$: resonance frequency

$Z_{11}, Z_{22}, Z_{12}, Z_{21}$: network impedances;

$A(\omega)$: open-loop gain;

$G(\omega)$: gain of amplifier

$W_j$: $j$-th working point;

$V_w$: voltage at working point;

$n_w$: number of periods;

$\Phi_w$: flux at working point;

$\Phi_f$: feedback flux

$V_n$: voltage noise;

$\Phi_n$: flux noise;

$V_f$: voltage of FLL;

$R_f$: feedback resistance;

$M_f$: mutual inductance of feedback coil;

$\partial V_s/\partial \Phi_a$: flux-to-voltage transfer-coefficient;

$R_{ftr}$: trans-impedance of dc-SQUID;

$V_{f\_FS}$: full-scale range of $V_f$;

$\Phi_{e\_FS}$: full-scale range of $\Phi_e$;

$R_{wire}$: wire resistance;

$SR_{max}(f)$: maximum slew rate;

$GBP(f)$: GBP of amplifier;

$V_m$: voltage of preamplifier

$R_d$: dynamic resistance of dc-SQUID;

$R_n$: noise-impedance of preamplifier;

$M_I$: mutual inductance for current feedback;

$M_V$: mutual inductance for voltage feedback;

$R_I$: impedance of current feedback;

$R_V$: impedance of voltage feedback;

$\Phi_{th}$: threshold value of flux input;

$\Phi_{in}$: flux input;

$j_x$: equivalent current that generates target magnetic fields;

$i_S$: average value of $i_{cir}$;

# 1. Introduction

After the discoveries of Meissner effect [1] and Josephson effect [2], various superconducting devices and circuits are developed for both analog and digital applications [3]; for example, flux transformers and superconducting quantum interference devices (SQUIDs) are devices for designing magnetic-field sensors [4], [5]; single-flux-quantum (SFQ) logics [6] are applied for power-efficient computing [7], and superconducting qubits [8] are developed for quantum computing [9]. Superconductive elements and devices [10-16] for the design of magnetic field sensors are illustrated in Fig. 1. Those devices are networks composed of resistor-inductor-capacitor (RLC) elements and two superconductor elements: superconductor wire and Josephson junction, as shown in Fig. 1(a) and (b). For instance, a flux-transformer is a closed superconducting loop of superconductor wires, as shown in Fig. 1(c); a superconducting loop inserted with one Josephson junction forms the radio-frequency (rf) SQUID, and it inserted with two Josephson junctions is called the direct-current SQUID, as shown in Fig. 1(d) and (e) respectively.

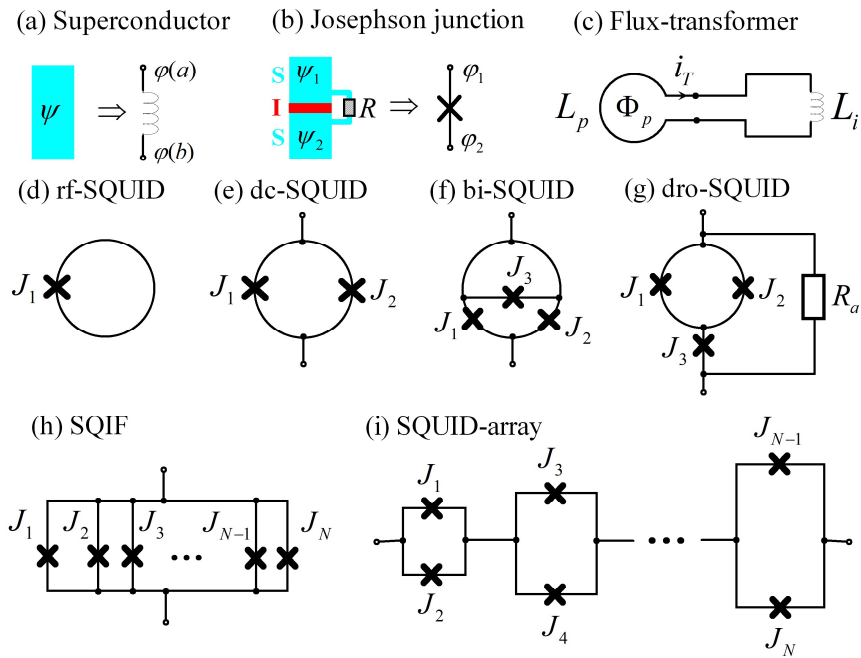

Fig. 1. Superconductor elements and devices: (a) superconductor wire; (b) resistively-shunted Josephson junction; (c) flux-transformer; (d) radio-frequency (rf) SQUID [10]; (e) dc-SQUID [11], [12]; (f) bi-SQUID [13]; (g) dro-SQUID [14]; (h) SQIF [15]; (i) SQUID-array [16].

Flux transformers are flux-to-current convertors, and SQUIDs are the most sensitive flux-to-voltage convertors; they are combined to develop many advanced magnetic field measurement systems successfully applied for biomagnetism [17], [18] and geophysical applications [19], such as magnetocardiogram (MCG), magnetoencephalogram (MEG), and full-tensor magnetic gradiometers (FTMG). In those SQUID sensors, the flux-transformer and the dc-SQUID are the flux-input front ends for flux-to-voltage conversions, while the rest of circuits is still implemented by room-temperature semiconductor amplifiers [20]. However, electronic engineers know well the semiconductor electronics, but not the superconductor electronics, in the design of SQUID systems, for the lack of superconductor physics.

In this article, we introduce the principles of SQUID sensors in the way that can be easily understood by the electronics engineers. We interpret the basic principles of superconductor elements and derive the general network equation of superconductor integrated circuits to reveal the working mechanism inside SQUIDs; We find the general analytical expression of dc SQUIDs, and present a magnetic field-effect transistor (MFET) concept to describe the function of SQUIDs in the design of the flux- or current-based analog circuits. It is shown that SQUID-based analog circuits are dual to the analog circuits based on the metal-oxide semiconductor field-effect transistors (MOSFETs), in both the design concepts and working principles.

## 2. Overview of SQUID electronics

### 2.1 Basic concepts

The overview of SQUID electronics, ranged from basic superconductive elements to sensor system, is illustrated in Fig. 2, where the concepts and objects for the design of SQUID systems are described as follows:

1) **Superconductor wire**: it is a strip of superconductor, which is not only a perfect conductor, but also a magnetic field coupler. its macroscopic phase difference at two terminals records the integral of magnetic vector potentials [21].

2) **Resistively-shunted Josephson junction**: it is a nonlinear phase-difference creator driven by the current; its equivalent circuit is described by the so-called Resistively-capacitively-shunted junction (RCSJ) model [22].

3) **Flux-transformer**: it is a superconducting loop connected with pure superconductor wires, which converts the flux coupled by the loop proportionally into superconducting current circulating inside the loop [5].

4) **Magnetic-field-effect transistor (MFET)** [23]: it is a kind of nonlinear device which resistance is modified by the applied magnetic fields; dc-SQUID is a kind of MFET; its resistance is modulated by the flux coupled by the superconducting loop. If the flux inputs of MFETs are applied through input coils, the combination of a MFET and the input coils is the current-inputted MFET (CI-MFET).

5) **Flux-feedback operational amplifier (FF-OPA)** [24]: it is a kind of operational amplifier simply realized by using a MFET as the flux-input front-end before a room-temperature operational amplifiers, where the MFET is used to implement the flux-to-voltage conversion.

6) **Flux-locked loop** [25]: it is exactly a flux-follower implemented by the FF-OPA working in closed-loop, similar as the voltage-follower implemented by the semiconductor voltage-feedback operational amplifier (VF-OPA).

7) **SQUID magnetometer or gradiometer**: it is realized with a FLL tightly coupled

to a flux-transformer. constitute, depending on the style of the pick-up coil.

8) **Trans-impedance amplifier**: a FLL integrated with an input coil works as a SQUID-based trans-impedance amplifier (TIA) [26]. A SQUID sensor is a SQUID-based TIA connected with a pick-up coil; magnetometers and gradiometers are only different in the style of pick-up coils.

9) **Multi-channel SQUID system**: it is a muti-channel current-sensing system based on SQUID sensors. For example, the SQUID-based magnetocardiography (MCG) and magnetoencephalography (MEG) systems use muti-channel SQUID sensors to pick up and read out the magnetic fields generated by the target current flows; they record the real-time signals of the target magnetic fields, and finally study the target through solving the inverse problem [17].

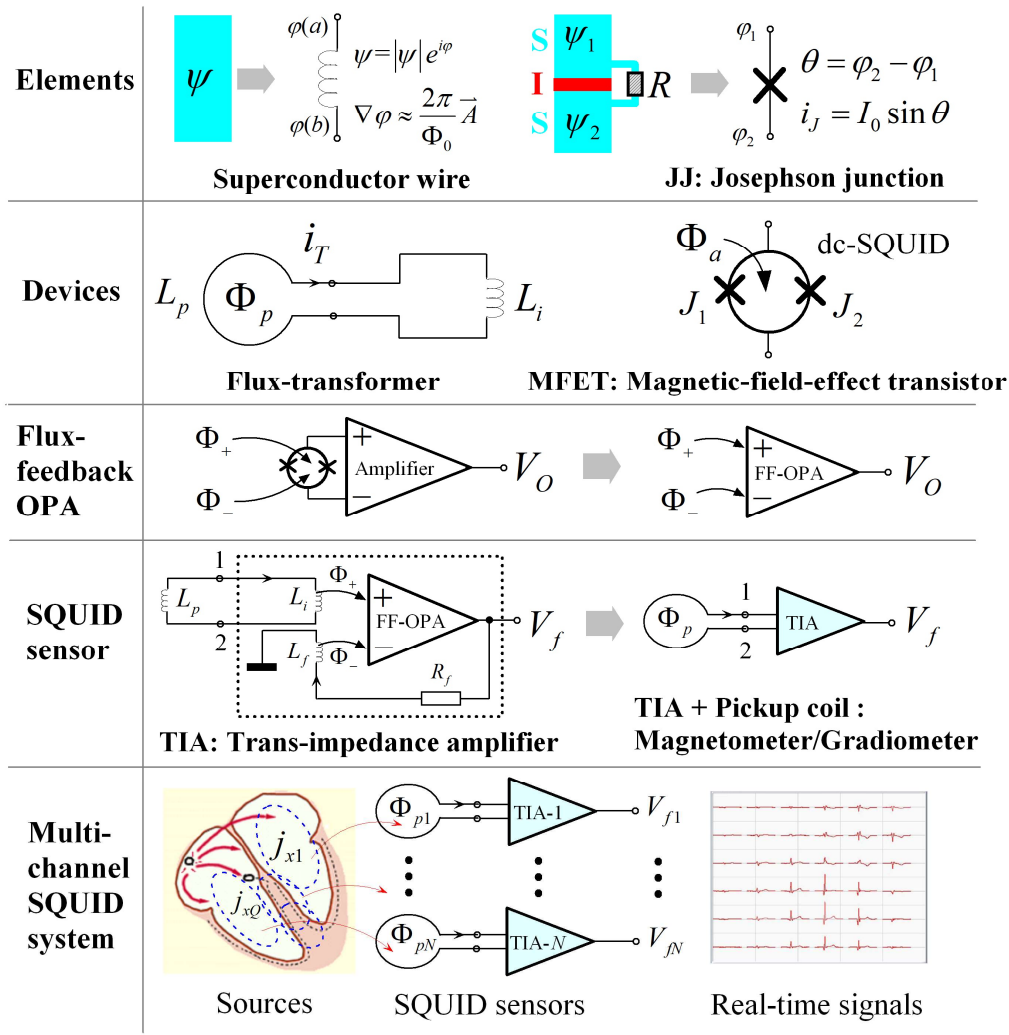

Fig. 2. Overview of SQUID-based analog circuits and systems.

The concepts of SQUID electronics above bring us an insight that, SQUID sensors are exactly the analog 'integrated' circuits of superconducting devices and semiconductor amplifiers, although SQUIDs and room-temperature amplifiers have to be installed separately for different working temperatures.

## 2.2 Levels of abstraction

There are four levels of abstraction in the design of a SQUID sensor systems, as summarized in Fig. 3:

1) At **device level**, we use two superconductor elements, superconducting wire and Josephson junction, to design flux-transformers and SQUID-based MFETs. The circuit analysis methods for resistor-inductor-capacitor (RLC) networks are also suitable for the superconducting devices, as long as we use the macroscopic quantum phases as variables to replace the voltages.

2) At **circuit level**, FF-OPAs are the analog integrated circuits of SQUID-based MFETs and room-temperature amplifiers; they use fluxes as inputs, similar to the semiconductor OPAs, in the design of analog circuits and systems.

3) At **module level**, SQUID sensors are the combination of a pick-up coil and a SQUID-based TIA; they combine the functions of the flux-transformer and the closed-loop FF-OPA.

4) At **system level**, a multi-channel SQUID system is a group of SQUID sensors configured with different pick-up coils; it is a contactless current sensing system, where, the pick-up coils sense the currents flowing inside the object under test, and linearly convert them into voltages through SQUID TIAs.

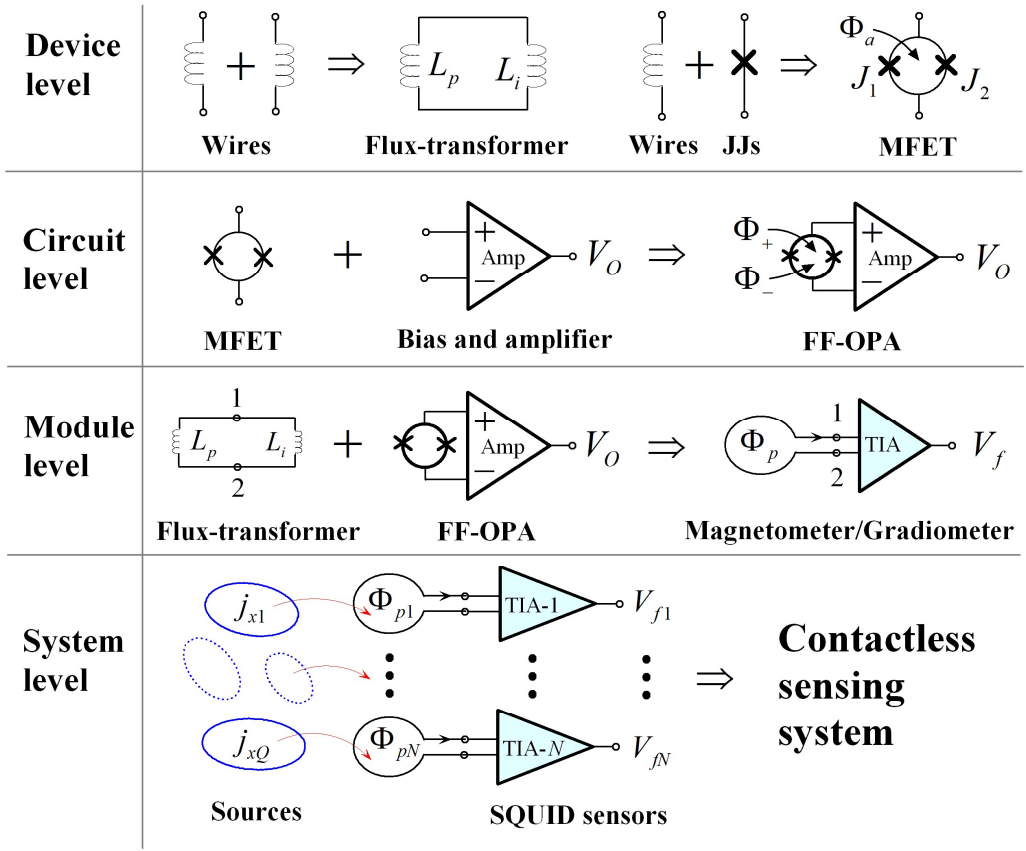

Fig. 3. Levels of abstraction for SQUID-based analog circuits and systems.

## 2.3 Duality between MFET and MOSFET

Fig. 4 exhibits the charge-flux duality between MFET and MOSFET electronics. MOSFETs and their application circuits and systems are electric-field or charge oriented [27], while SQUID-based MFETs and their circuits and systems are magnetic-filed oriented; two devices use similar schemes in the design of analog circuits and systems.

The comparison between SQUID-based MFET and MOSFET electronics is summarized in Table 1.

Table 1. Comparison between SQUID and MOSFET electronics

|  | SQUID electronics | MOSFET electronics |
|---|---|---|
| Physics | Superconductor physics | Semiconductor physics |
| Circuit theory | Flux-oriented | Charge-oriented |
| Elements | Josephson junctions and RLC elements | PN junctions and RLC elements |
| Device level | Magnetic FET | Electric FET |
| Circuit level | Flux/current Feedback OPA | Charge/voltage Feedback OPA |
| Module level | Flux/current follower | Charge/voltage follower |
| System level | Flux/current sensing system | Charge/voltage detecting system |

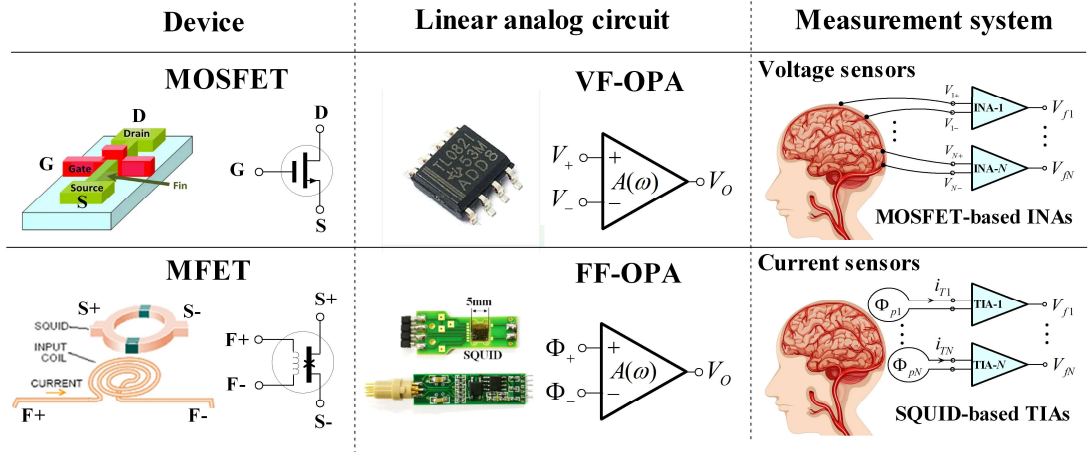

Fig. 4. Charge-flux duality between MOSFET and SQUID electronics.

## 2.4 Circuit theory for SQUID electronics

Four variables are used in the analysis of electric circuits [28], as shown in Fig. 5. The charge $Q$ and flux $\Phi$ are the entities of electric and magnetic fields; their flow rates are current $i$ and $v$, respectively. The principles of using circuit variables are:
1) Magnetic field-oriented circuits are the flux-current systems; they are usually analyzed with $\Phi$ and $i$ as variables;
2) Electric field-oriented circuits are the charge-voltage systems; they are usually analyzed with $Q$ and $v$ as variables.

SQUIDs and their application circuits are magnetic-filed oriented, and are analyzed as flux-current circuit systems. Details about circuit variables and analysis methods are seen in the reference [29].

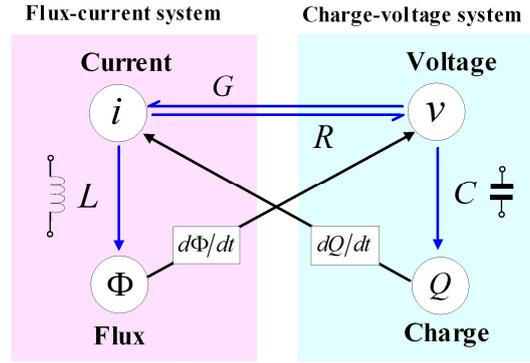

Fig. 5. Variables and elements in the electric circuit analyses.

## 3. Principles of superconductor elements

### *3.1 Superconductor elements*

Superconductor circuits are superconducting RLC networks, in which superconductor wire and Josephson junction shown in Fig. 6 are the two unique elements that make superconductor circuits different from normal RLC circuits.

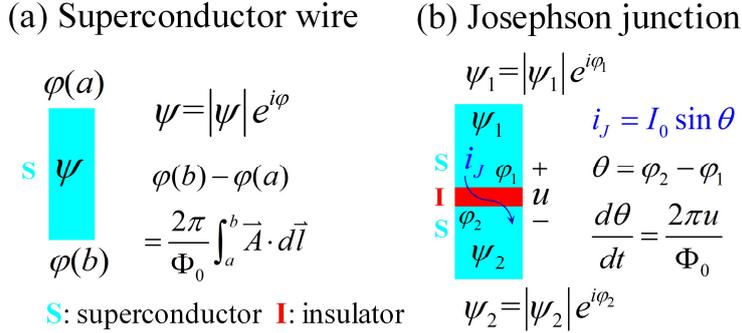

Fig. 6. Superconductor elements: (a) Superconductor wire, and (b) Josephson junction.

Superconductor wires achieve two properties:
1) They are the perfect conductors without resistance, similar with the wires in the conventional circuit graphs;
2) They are the magnetic-field couplers, which sense the magnetic field with their charge carriers [21].

Cooper pairs are the charge carriers of superconductor wires; their condense state is described by a macroscopic wave function, $\psi = |\psi|\, e^{i\varphi}$, as shown in Fig. 6(a). According to the Ginzburg-Landau theory [21], the quantum phase $\varphi$ will vary along the wire with a gradient proportional to $A$; the phase difference between terminal-*a* and terminal-*b* of superconductor wire is

$$\varphi(b) - \varphi(a) = \frac{2\pi}{\Phi_0} \int_a^b \vec{A} \cdot d\vec{l} \qquad (1)$$

where $\Phi_0$ is the flux quantum, $\Phi_0 = h/2e = 2.07 \times 10^{-15}$ Wb.

Josephson junctions have a superconductor-insulator-superconductor (SIS) structure, as illustrated in Fig. 6(b), where $\varphi_1$ is the phase at the '+' terminal and $\varphi_2$ is the one at the '–' terminal; the two-terminal phase difference is $\theta$, and $\theta = \varphi_2 - \varphi_1$, which changes with the voltage $u$. Josephson current $i_J$ is tunnelling through the insulator, which relation with the two-terminal phase and voltage is

$$i_J = I_0 \sin\theta$$
$$\begin{cases} \theta = \varphi_2 - \varphi_1 \\ \dfrac{d\theta}{dt} = \dfrac{2\pi u}{\Phi_0} \end{cases} \tag{2}$$

where $I_0$ is the critical current of the Josephson junction.

Josephson junctions implement two functions in superconducting circuits:
1) They are phase-difference generators, which create a phase gap between two superconductors.
2) Their Josephson currents are voltage-controlled oscillators (VCOs) [30-32], which are oscillating with a fundamental frequency $\omega_{rf}$ controlled by the average voltage $V_s$ at the two terminals.

The VCO function of Josephson currents is described as follows:

$$i_J = I_0 \sin\theta = I_0 \sin\left(\omega_{rf} t + \theta(0) + \Delta\theta\right)$$
$$\begin{cases} \omega_{rf} = \dfrac{2\pi V_s}{\Phi_0}; |\Delta\theta| < C_\theta \\ V_s = \dfrac{1}{T}\int_0^T u \cdot dt; T \gg \dfrac{2\pi}{\omega_{rf}} \end{cases} \tag{3}$$

where $C_\theta$ is a bounded constant for the $\Delta\theta$, and $\omega_{rf}$ is the angular frequency controlled by the average voltage $V_s$, with a voltage-to-frequency ratio of 0.438 GHz/μV.

### 3.2 Closed-loop law

The closed loops connected by two kinds of superconductor elements will comply with the fluxoid quantization law (FQL) [21]. A superconducting loop connected by $N$ Josephson junctions ($N$ is an integer) and $N$ superconductor is illustrated in Fig. 7, where each wire has a phase difference generated by the magnetic vector potential, and each gap between two wires has a phase difference created by the Josephson junction. Those phase differences will meet the uniqueness of the macroscopic wave function inside wires as

$$\sum_{i=1}^{N}(\varphi_i(a_i) - \varphi_i(b_i)) + \sum_{i=1}^{N}\sigma_i \theta_i = 2n\pi$$
$$n \in \mathbb{Z}; \sigma_i = \begin{cases} +1; & \text{if } i_{cir} \text{ enters '+' of } \theta_i \\ -1; & \text{if } i_{cir} \text{ enters '–' of } \theta_i \end{cases} \tag{4}$$

The factor $\sigma_i$ exhibits the polarity of $\theta_i$ inside the loop, with the direction of loop current $i_{cir}$ as the reference. For instance, in Fig. 7, $\sigma_1$ is 1, since $i_{cir}$ enters the '+' terminal of $\theta_1$, while $\sigma_2$ is –1, since $i_{cir}$ enters the '–' terminal of $\theta_2$.

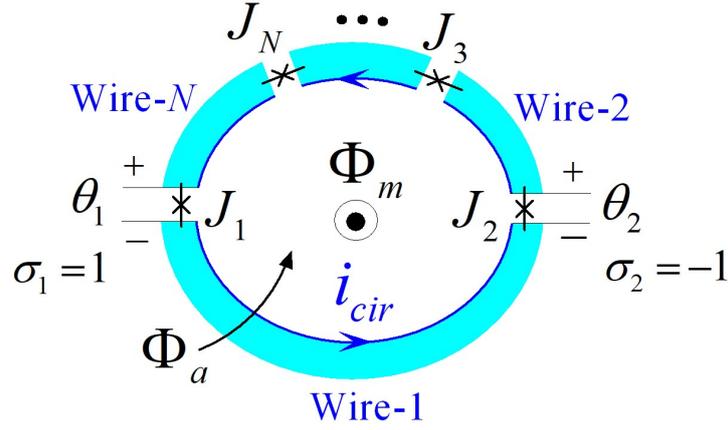

Fig. 7. Superconducting loop connected by *N* Josephson junctions

According to (1), we can sum the phase differences of *N* superconductive wires, and find that

$$\sum_{i=1}^{N}(\varphi_i(a_i) - \varphi_i(b_i)) = \frac{2\pi}{\Phi_0}\sum_{i=1}^{N}\left(\int_{a_i}^{b_i}\vec{A}\cdot d\vec{l}\right) \approx \frac{2\pi}{\Phi_0}\oint_l \vec{A}\cdot d\vec{l} = \frac{2\pi}{\Phi_0}\Phi_m \qquad (5)$$

Therefore, we can derive the so-called fluxoid-quantization law (FQL) as

$$\Phi_m + \frac{\Phi_0}{2\pi}\sum_{i=1}^{N}\sigma_i\theta_i = n\Phi_0 \qquad (6)$$

The derivative of FQL with respect to time is the Faraday's law of electromagnetic induction and the Kircholff's voltage law (KVL):

$$\frac{d\Phi_m}{dt} + \sum_{i=1}^{N}\sigma_i u_i = 0 \qquad (7)$$

In other words, the FQL is the integral form of the Faraday's law and KVL, in which the constant is specified as $n\Phi_0$. Therefore, the closed-loop law in (6) is general for both superconducting and normal loops.

In a loop with a self-inductance $L_m$, the flux $\Phi_m$ coupled by the loop is applied by the external flux $\Phi_a$ and generated by the loop current $i_{cir}$, thus,

$$\Phi_m = L_m i_{cir} - \Phi_a \qquad (8)$$

where the self-induced flux is assumed to cancel the applied flux $\Phi_a$.

# 4. Flux transformer

## *4.1 Principle of flux-transformer*

A Flux-transformer is a superconducting loop purely connected by superconductor wires, as illustrated in Fig. 8, where there is a superconducting current, namely $i_T$, flowing through a pick-up coil $L_p$ and an input coil $L_i$. According to the FQL, the total flux coupled in the loop must be quantized as

$$\Phi_m = n_T\Phi_0 \qquad (9)$$

where, $n_T$ is the number of flux quanta trapped in the loop.

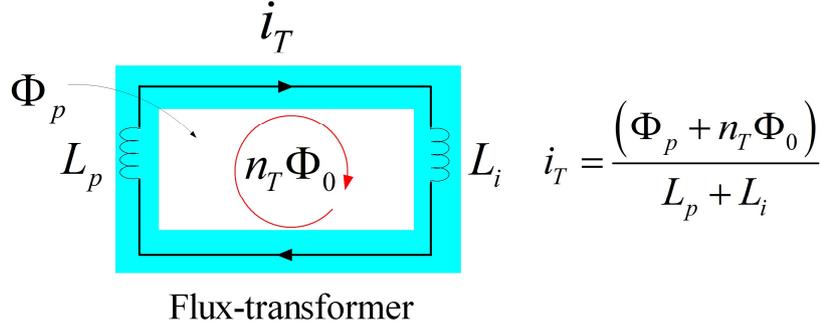

Fig. 8. Flux-transformer consists of a pick-up coil and an input coil.

The flux-transformer will turn the flux $\Phi_p$ picked-up by the pick-up coil linearly into loop current $i_T$; the function of $i_T$ will be calculated by $\Phi_p$ as

$$\Phi_m = (L_p + L_i)i_T - \Phi_p \Rightarrow i_T = \frac{\Phi_p + n_T \Phi_0}{L_p + L_i} \tag{10}$$

Accordingly, flux transformers are linear flux-to-current convertors.

### 4.2 Styles of pick-up coils

Pick-up coils have various styles to extract the components of magnetic fields [4], [5], as illustrated in Fig. 9. The pick-up coil for magnetometers is winding along a planar surface, as shown in Fig. 9(a); the first-order gradiometer has two planar surfaces, as shown in Fig. 9(b); the second-order gradiometer have four planar surfaces, as exhibited in Fig. 9(c). The flexible configuration of pick-up coils for different targets is one of the advantages of SQUID sensors.

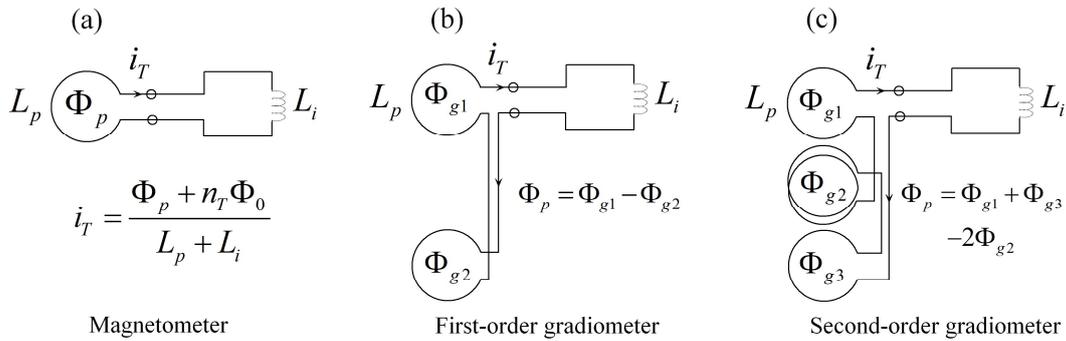

Fig. 9. Flux transformers: (a) pick-up coil is a magnetometer; (b) pick-up coil is a first-order gradiometer; (c) pick-up coil is a second-order gradiometer.

## 5. Magnetic field-effect transistor

### 5.1 Structure of dc-SQUID

A dc-SQUID is a superconducting loop inserted with two Josephson junctions, as shown in Fig. 10(a), where two terminals of Josephson junctions are led to a bias current source $I_b$, and a non-zero average voltage $V_s$ will be measured when the bias current is larger than the critical current of Josephson junctions. The equivalent circuit of the dc-SQUID is a RLC network driven by two Josephson currents, as shown Fig. 10(b), where $i_{n1}$ and $i_{n2}$ are noise currents. Each Josephson junction is treated as a RCSJ [22], and is symbolized as shown in Fig. 10(c); accordingly, the dc-SQUID is represented with a two-terminal device as shown in Fig. 10(d).

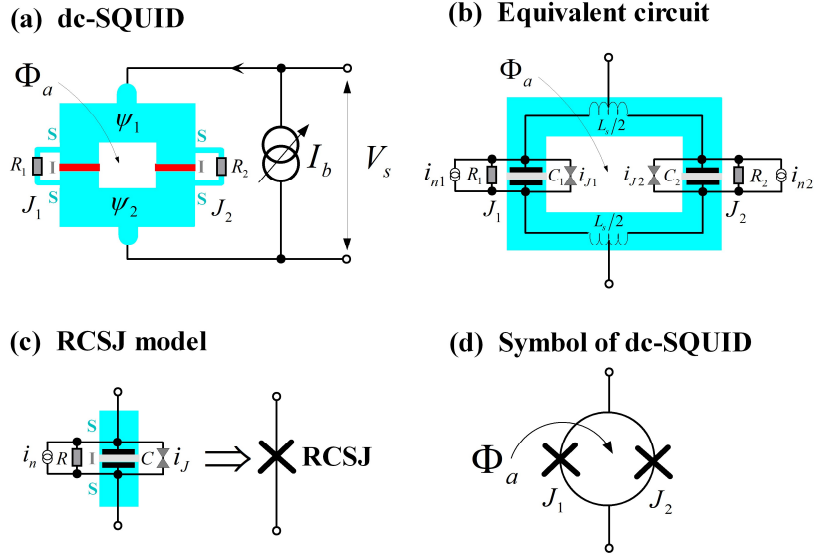

Fig. 10. (a) dc-SQUID under current bias. (b) Equivalent circuit of dc-SQUID. (c) RCSJ model. (d) Symbol of dc-SQUID

## 5.2 Circuit equations of dc-SQUID

The dc-SQUID loop contains a superconducting current $i_{cir}$, as defined in Fig. 11(a). The $i_1$ and $i_2$ flowing through two Josephson junctions are supplied by the $i_{cir}$ and bias current $I_b$; therefore, the dc-SQUID can be redrawn as shown in Fig. 11(b), where each Josephson junction is separately biased with $I_b/2$.

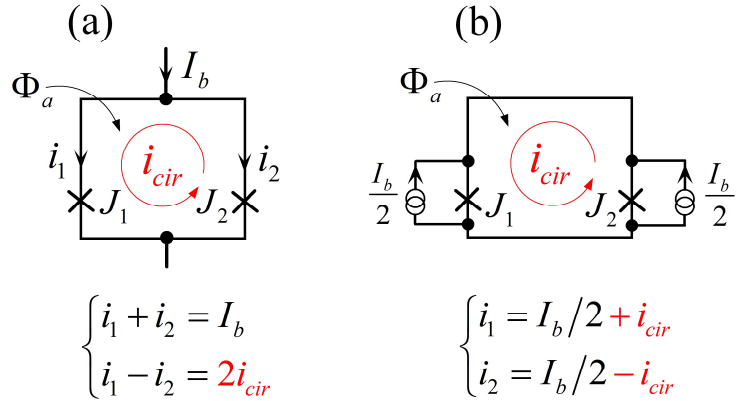

Fig. 11. (a) Superconducting current circulating in dc-SQUID loop. (b) Equivalent circuit of dc-SQUID, where two Josephson junctions are separately biased.

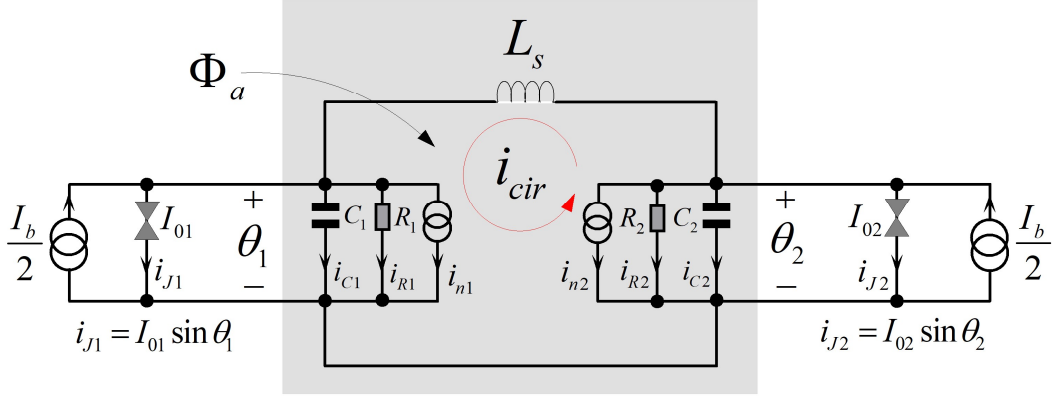

Fig. 12. Two-port dc-SQUID network driven by Josephson currents [23].

The dc-SQUID is equivalent to a two-port network equivalent circuit [23], as shown in Fig. 12. The two-port RLC network is driven by two Josephson currents and two bias currents. Josephson currents, $i_{J1}$ and $i_{J2}$, are the phase-controlled current sources; the $I_b$ and $\Phi_a$ are the given inputs, while the $\theta_1$ and $\theta_2$ are two outputs. The circuit equations of the dc-SQUID are derived as follows:

First, the FQL in the SQUID loop is written as

$$L_s i_{cir} - \Phi_a + \frac{\Phi_0}{2\pi}(\theta_1 - \theta_2) = 0 \tag{11}$$

Second, two current-phase relations of Josephson junctions are

$$\begin{cases} \dfrac{\Phi_0 C_1}{2\pi}\theta_1'' + \dfrac{\Phi_0}{2\pi R_1}\theta_1' + I_{01}\sin\theta_1 + i_{n1} = \dfrac{I_b}{2} + i_{cir} \\ \dfrac{\Phi_0 C_2}{2\pi}\theta_2'' + \dfrac{\Phi_0}{2\pi R_2}\theta_2' + I_{02}\sin\theta_2 + i_{n2} = \dfrac{I_b}{2} - i_{cir} \\ \theta_1'' = \dfrac{d^2\theta_1}{dt^2}; \theta_2'' = \dfrac{d^2\theta_2}{dt^2}; \theta_1' = \dfrac{d\theta_1}{dt}; \theta_2' = \dfrac{d\theta_2}{dt} \end{cases} \tag{12}$$

### 5.3 Dynamics inside dc-SQUID

The circuit equations of dc-SQUID in (11) and (12) are exhibited by a system diagram shown in Fig. 13 [33]. The module colored in grey is a second-order linear system modified by two nonlinear phase-to-current feedback functions, $i_{J1}$ and $i_{J2}$. The dynamics of this system are described as follows:

1) In zero-voltage state, $\theta_1$ and $\theta_2$ are stable in a certain value that enable $i_{J1}$ and $i_{J2}$ to cancel the $i_1$ and $i_2$ supplied by $I_b$ and $i_{cir}$.
2) In voltage state, $\theta_1$ and $\theta_2$ will keep increasing because $i_{J1}$ and $i_{J2}$ are less than $i_1$ and $i_2$; the average increasing speed of $\theta_1$ and $\theta_2$ is proportional to the measured voltage $V_s$ of dc-SQUID.
3) The external $\Phi_a$ is turned into $i_{cir}$ to adjust the bias currents of two Josephson junctions, and thereby modulates the average speed of $\theta_1$ and $\theta_2$; this is how the flux-modulated current-voltage characteristics of dc-SQUID are formed.

Therefore, under the same average voltage $V_s$, two Josephson currents are working as two VCOs, and are driving the two-port RLC network in oscillation state; the fundamental oscillation frequency $\omega_{rf}$ is defined in (3).

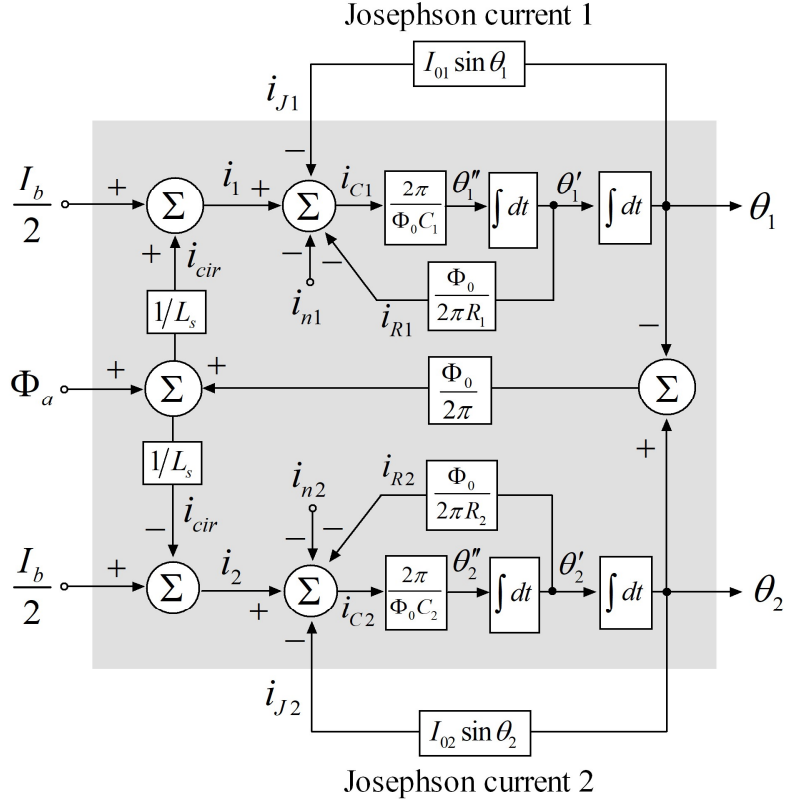

Fig. 13. System diagram of dc-SQUID [33].

## 5.4 Magnetic-flux-flow diagram of dc-SQUID

SQUIDs and single-flux-quantum (SFQ) gates, as well as Josephson qubits, are all the superconducting RLC networks driven by additional Josephson currents, and are unified with normal RLC circuits with a common network equation [34], [35]. The general network equation in matrix is

$$\mathbf{C}\frac{d^2\mathbf{\Phi}_{EL}}{dt^2}+\mathbf{G}\frac{d\mathbf{\Phi}_{EL}}{dt}+\mathbf{I_0}\sin\left(\frac{2\pi\mathbf{\Phi}_{EL}}{\Phi_0}\right)+\sigma^T\mathbf{L_m^{-1}}\sigma\mathbf{\Phi}_{EL} \qquad (13)$$
$$=\mathbf{i_b}-\mathbf{i_n}+\sigma^T\mathbf{L_m^{-1}}\left(\mathbf{Const}+\mathbf{\Phi_a}\right)$$

The variable vectors and parameter matrices for the dc-SQUID shown in Fig. 12 are

$$\begin{cases}\mathbf{\Phi}_{EL}=\begin{bmatrix}\Phi_{EL1} & \Phi_{EL2}\end{bmatrix}^T=\frac{\Phi_0}{2\pi}\begin{bmatrix}\theta_1 & \theta_2\end{bmatrix}^T;\mathbf{Const}+\mathbf{\Phi_a}=\begin{bmatrix}\Phi_a\end{bmatrix}\\ \mathbf{i_b}=\begin{bmatrix}\frac{I_b}{2} & \frac{I_b}{2}\end{bmatrix}^T;\mathbf{i_n}=\begin{bmatrix}i_{n1} & i_{n2}\end{bmatrix}^T;\sigma=\begin{bmatrix}1 & -1\end{bmatrix};\mathbf{L_m}=\begin{bmatrix}L_s\end{bmatrix}\\ \mathbf{C}=\begin{bmatrix}C_1 & 0\\ 0 & C_2\end{bmatrix};\mathbf{G}=\begin{bmatrix}1/R_1 & 0\\ 0 & 1/R_2\end{bmatrix};\mathbf{I_0}=\begin{bmatrix}I_{01} & 0\\ 0 & I_{02}\end{bmatrix}\end{cases} \qquad (14)$$

This general network equation is derived by a genal flux-based circuit theory [36]. In this theory, Josephson junction circuits can be described with a general network equation and simulated with a general system model; they can also be redrawn with magnetic-flux-flow (MFF) diagrams [35] to vividly depict flux-flow dynamics.

For example, the dc-SQUID shown in Fig. 14(a) is further exhibited with a magnetic-flux-flow diagram, as shown in Fig. 14(b). This MFF diagram reveals the flux-flow mechanism inside dc-SQUID as follows:

1) Two current-biased Josephson junctions are two magnetic-flux generators (MFGs), namely MFG-1 and MFG-2. In the conventional circuit diagram, two Josephson junctions are inserted in the SQUID loop, Loop-1, while they are connected to Loop-1 in the MFF diagram;
2) MFG-2 keep pumping flux quanta into Loop-1, while MFG-1 keep drawing the flux quanta out of Loop-1. The average flow-rate of the flux flow pumping in and out Loop-1, is exactly the average voltage $V_s$ that is measured at two terminals of MFG-1 and MFG-2.
3) The external flux $\Phi_a$ applied to Loo-1 is turned into an offset current applied to two Josephson junctions, according to (13); it will therefore alter the flow-rate, like a water-tap.

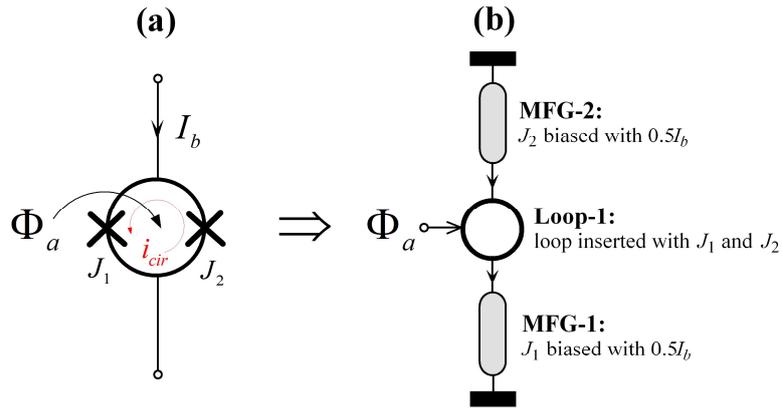

Fig. 14. (a) dc-SQUID and (b) its magnetic-flux-flow (MFF) diagram [36].

This MFF diagram vividly depicts the flux-flow dynamics inside dc SQUIDs working as MFETs. More details are seen in [35] [36].

### 5.5 Characteristics of dc-SQUID

The current-voltage characteristics of dc-SQUID are illustrated in Fig. 15. They are captured by measuring $V_s$ under various $\Phi_a$ and $i_b$, from the dc-SQUID shown in Fig. 15(a); They can also be numerically simulated with either the system model shown in Fig. 13 or the circuit equation in (13). The measured characteristics can be well agreed by the simulations [35], if the circuit model and parameters are properly set.

With $V_s$ and $i_b$ as the coordinates, the current-voltage curves under different $\Phi_a$ are shown in Fig. 15(b). Those curves are expressed with a function as

$$V_s = f(i_b, \Phi_a) = f(i_b, \Phi_a + n\Phi_0); n \in \mathbb{Z} \tag{15}$$

It is a periodical function for the flux input $\Phi_a$, in which, the period is $1\Phi_0$. By setting $i_b$ as $I_b$, we can observe a periodical flux-voltage characteristic, as illustrated in Fig. 15(c); this flux-voltage curve is defined with a function as

$$V_s = f_{V-\Phi}(\Phi_a) = f(I_b, \Phi_a) \tag{16}$$

Those functions achieve practical physical meanings, in the system model shown in Fig. 13 and the MFF diagram shown in Fig. 14. In the system model in Fig. 13, those functions describe how the average increasing speed of $\theta_1$ and $\theta_2$ is modulated by $i_b$ and $\Phi_a$; they also describe how the $i_b$ and $\Phi_a$ tune the average flow rate of the flux flow in MFG1 and MFG2, in the MFF diagram shown in Fig. 14.

In the current-voltage curves shown in Fig. 15(b), there is a LC-resonance point, at which the flux-modulation characteristic disappears; the voltage $V_{s0}$ indicates the resonance frequency, and can be analytically expressed with the RLC parameters inside the SQUID washer [33]. How the resonance frequency $V_s$ under the given $i_b$ and $\Phi_a$ is decided by circuit parameters inside the dc-SQUID can be explained by the analytical expression of the current-voltage function defined in (15).

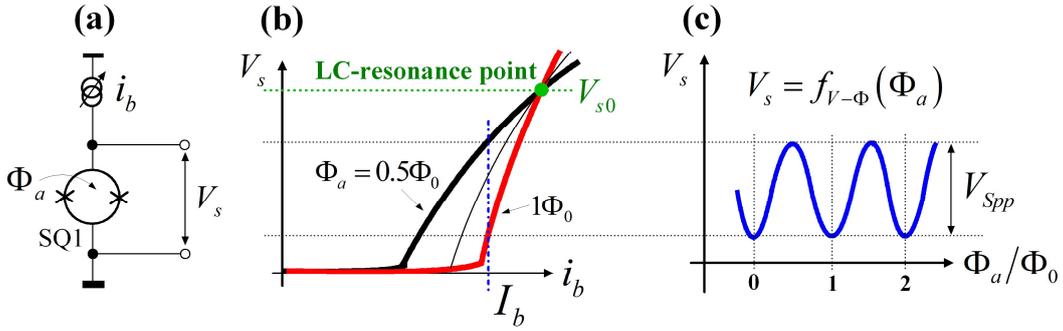

Fig. 15. Dc-SQUID and characteristics: (a) dc-SQUID under current bias; (b) flux-modulated current-voltage curves of dc-SQUID; (c) periodical flux-voltage curve of dc-SQUID.

### 5.6 Analytical expression of dc-SQUID

The analytical expression of (15) is difficult to be directly solved from the differential circuit equations that are varied with the equivalent circuit of dc SQUIDs [37-40]; it is still lacked in the long history of dc SQUIDs.

We proposed a general analytical expression of dc SQUIDs [23], in another form of

$$F(V_s, i_b, \Phi_a) = 0 \Rightarrow \begin{cases} i_b = f_I(V_s, \Delta\theta_{dc}) = \left(\dfrac{1}{Z_{11}(0)} + G_{cmm}\right) \cdot V_s \\ \Phi_a = f_\Phi(V_s, \Delta\theta_{dc}) = \dfrac{\Phi_0}{2\pi} \cdot \Delta\theta_{dc} + \Phi_{cir} + \eta L_s^* V_s + \alpha L_s^* i_b \end{cases} \tag{17}$$

where $i_b$ and $\Phi_a$ are the dependent variables of the resonance frequency $V_s$ and the average phase difference between $\theta_1$ and $\theta_2$, namely $\Delta\theta_{dc}$.

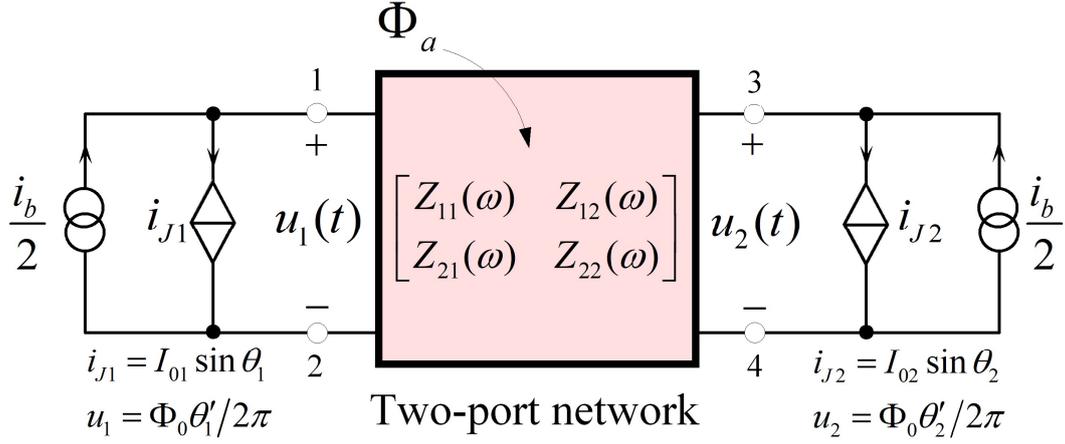

Fig. 16. A general two-port network model for dc SQUIDs [23].

This general analytical expression is derived by using a general two-port network model of dc SQUIDs, as shown in Fig. 16. The model treats any dc-SQUID circuit as a two-port network driven by the $i_b$ and two Josephson currents. This two-port network achieve flux-modulated current-voltage characteristics, based on two principles:
1) Under a nonzero voltage $V_s$, two Josephson currents are VCOs, they generate both dc and ac currents into two-port network; the dc component of Josephson currents is proportional to the total power transferred by the ac components, due to the zero-power consumption principle of Josephson currents.
2) In response to $i_b$ and two Josephson currents, the two-port network creates both the $V_s$ and ac voltages at two ports. In frequency domain, the two-port network implements its transfer-function with only four network impedances, such as $Z_{11}$, $Z_{22}$, $Z_{12}$, and $Z_{21}$ ($Z_{12} = Z_{21}$ for reciprocity principle).

The $G_{cmm}$ in (17) is the equivalent conductance of the dc component of two Josephson currents, we find the analytical expressions of $G_{cmm}$ with frequency domain analyses in [23]. Based on the expression of $G_{cmm}$, a dc-SQUID is equivalent to a flux-controlled nonlinear resistor in parallel with the $Z_{11}(0)$, as illustrated in Fig. 17.
1) A bare-washer dc-SQUID shown in Fig. 17(a) is equivalent to two resistors in parallel, as shown in Fig. 17(b).
2) A dc-SQUID coupled with one input coil and the equivalent circuit are exhibited in Fig. 17(c) and (d), where $Z_{i1}$ is the impedance at the terminals of the input coil, and is involved in four impedances, since the input coil is a part of the two-port network.
3) A dc-SQUID coupled with two input coils is modeled as shown in Fig. 17(e) and (f), where both $Z_{i1}$ and $Z_{i2}$ will affect the impedances of the two-port network, and thereby modify the current-voltage characteristics [41-44].

Details of the derivation of $G_{cmm}$ are seen in the reference [23]; it is shown that the $G_{cmm}$ is the projection of the network impedances, and the real-part of impedance $Z_{12}$, Re($Z_{12}$), decides the flux-modulation depth of current-voltage characteristics; the reason why there will be the crossing of current-voltage curves in Fig. 15(b) is that the real-part of $Z_{12}$ is zero at the LC-resonance point.

Accordingly, for a symmetric bare-washer dc-SQUID, we can derive that the $V_{s0}$ at the LC-resonance point is decided by the loop inductance and the capacitance of one Josephson junction [33], namely

$$\frac{V_{s0}}{\Phi_0} = \frac{1}{2\pi\sqrt{L_s C_j}}; C_j = C_1 = C_2 \tag{18}$$

To be notice that, $V_{s0}/\Phi_0$ is not the resonance frequency of the SQUID loop composed of two junction capacitances which resonance frequency is $1/\pi\sqrt{(2L_s C_j)}$. More details are seen in [23] [33].

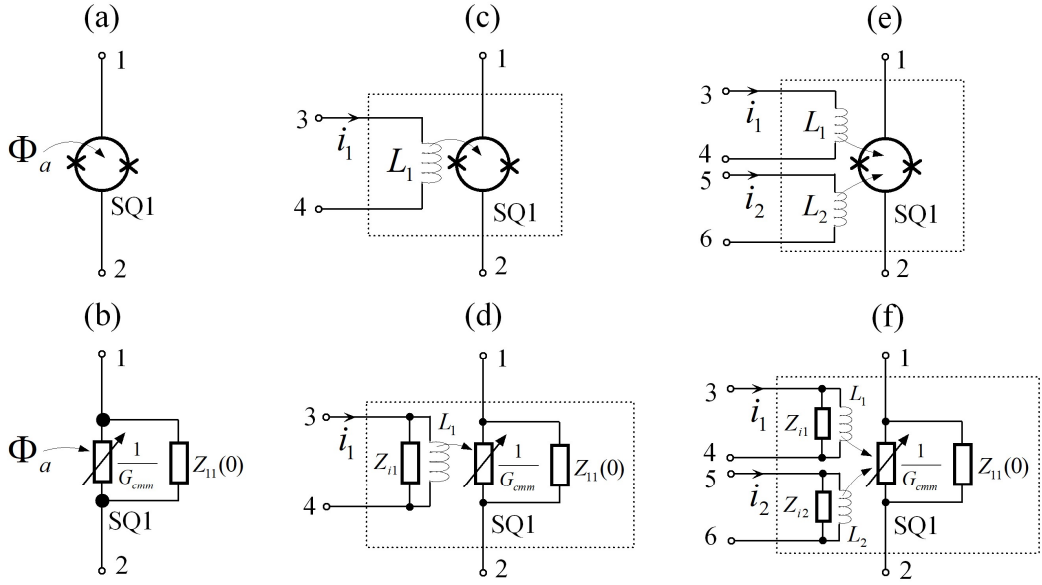

Fig. 17. Nonlinear resistance of SQUID-based MFET [23]: (a) bare-washer dc-SQUID, and (b) its nonlinear resistor model; (c) CI-MFET with one input coil, and (d) the equivalent circuit; (e) CI-MFET with two input coils, and (f) the equivalent circuit.

The general analytical expression of dc-SQUID in (17) will have two practical applications:
1) It can be used to characterize circuit parameters of a dc-SQUID with the measured current-voltage characteristics.
2) It can be used to predict the current-voltage characteristics of a dc-SQUID with the network impedances extracted from the circuit layout.

# 6. Flux-feedback operational amplifier

## 6.1 Concept of FF-OPA

Operational amplifiers (OPAs) are the general scheme to design linear analog circuits using nonlinear devices [45]. The voltage-feedback OPA (VF-OPA) and the flux-feedback OPA (FF-OPA) are illustrated in Fig. 18. With a three-terminal VF-OPA shown in Fig. 18(a), which open-loop-gain $A(\omega)$ far larger than 1, $A(\omega)\gg1$, a linear voltage follower is implemented by feeding the voltage output back to the inverting voltage input, as shown in Fig. 18(b). Similarly, with a three-terminal FF-OPA shown in Fig. 18(c), a linear flux-follower is realized by feeding the voltage output to the inverting flux input through a circuit consisting of a resistor and a feedback coil in serial, as illustrated in Fig. 18(d).

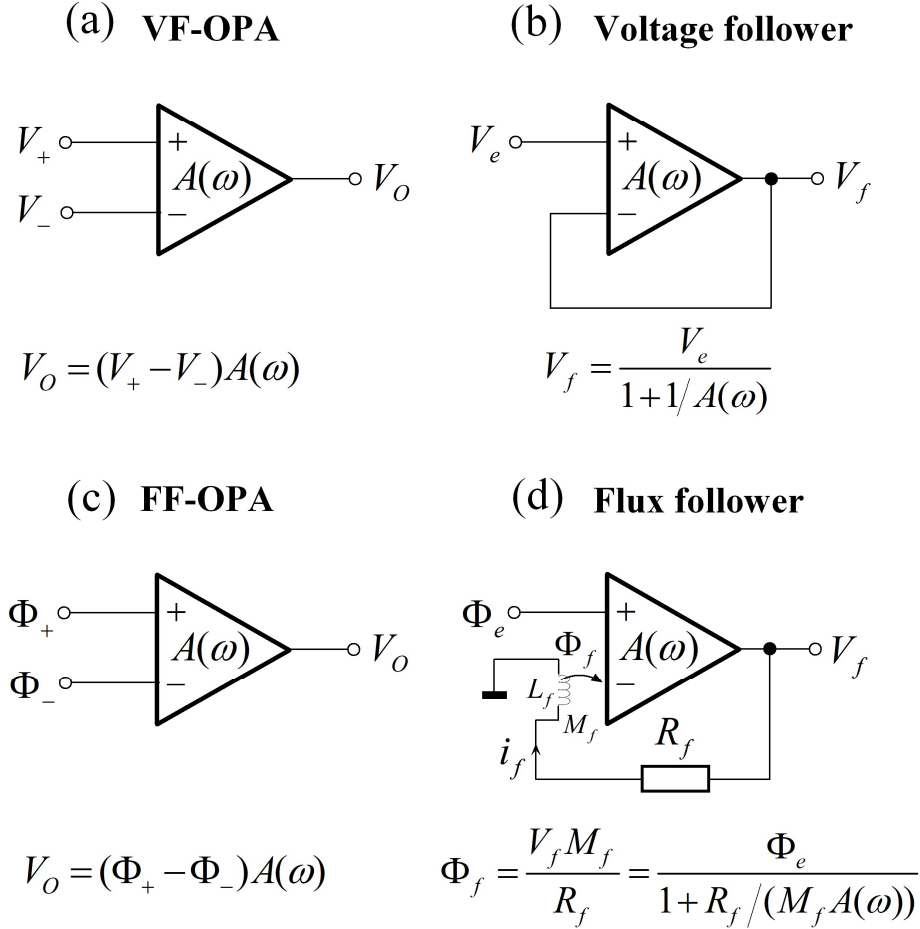

Fig. 18. (a) Voltage-feedback operational amplifier (VF-OPA) and its (b) voltage-follower circuit; (c) Flux-feedback operational amplifier (FF-OPA) and (d) its flux-follower circuit.

### 6.2 SQUID-based FF-OPA

A simple SQUID-based FF-OPA is shown in Fig. 19(a), where the dc-SQUID is simply a MFET used to sense the positive and negative flux inputs. The dc-SQUID is biased with a current source $I_b$ and connected to the non-inverting input of the room-temperature amplifier U1. The current source $I_r$ is applied on a resistor $R_r$ to generate a voltage at the inverting input of U1.

The flux-flower based on the negative feedback mechanism is accordingly implemented with the SQUID-based FF-OPA, as shown in Fig. 19(b), It is exactly the well-known flux-locked loop (FLL), which generates $\Phi_f$ to cancel the flux input $\Phi_e$ and keep the difference between $V_+$ and $V_-$ being zero, $V_+ = V_-$.

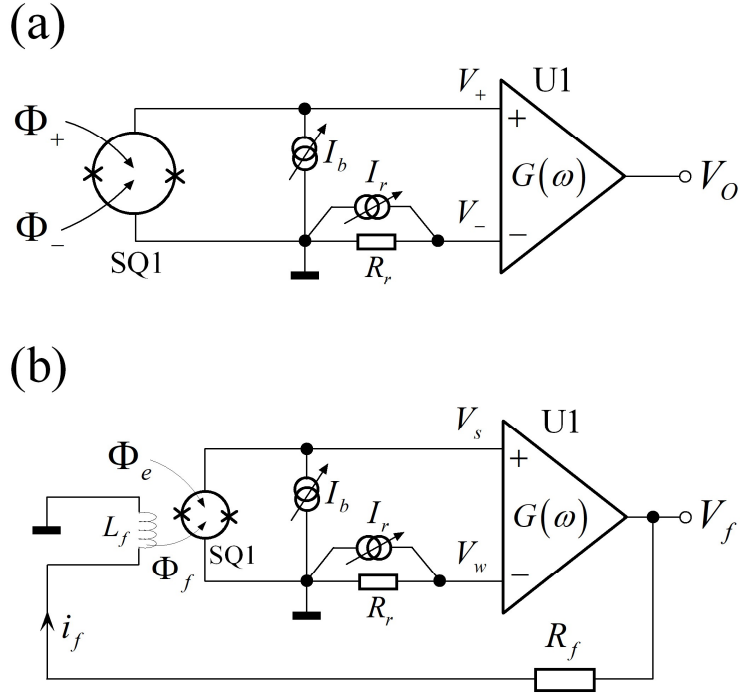

Fig. 19. (a) SQUID-based FF-OPA, (b) FLL implemented by a flux-follower [24].

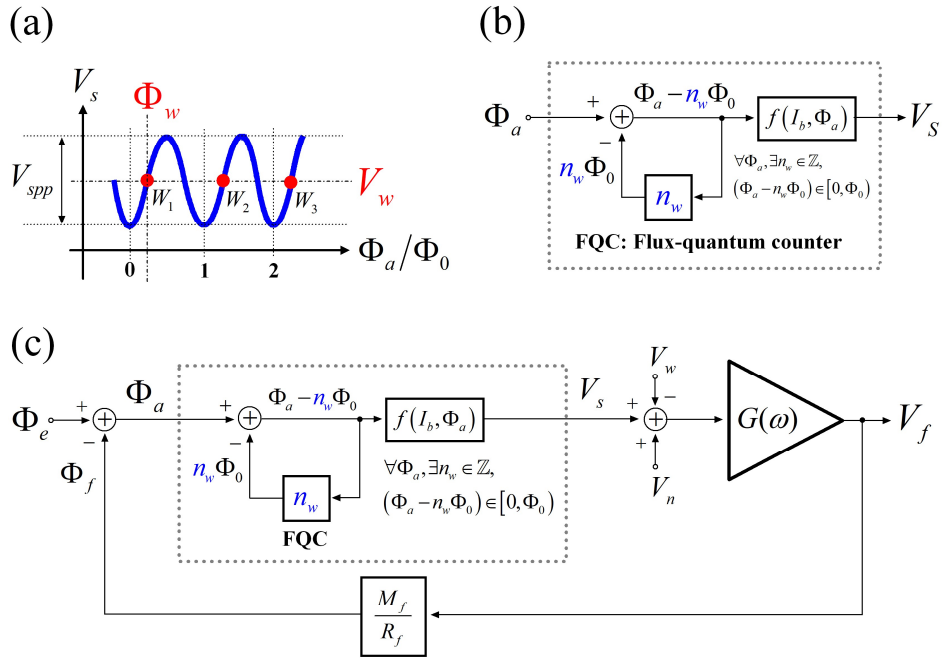

Fig. 20. (a) flux-voltage characteristic of dc-SQUID; (b) transfer function of dc-SQUID; (c) transfer function diagram of SQUID-based FLL [26].

Assuming that the input impedances of the U1 are far larger than the SQUID and $R_r$, and are ignorable, $V_+ = V_s$, $V_- = V_w$ and $V_w = I_r R_r$; the SQUID output $V_s$ will be locked at $V_w$ dynamically. The points, such as, $W_1$, $W_2$, and $W_3$, are the working points of the FLL periodically located in the positive slopes of the current-voltage characteristic, as illustrated in Fig. 20 (a). Since the $j$-th working point $W_j$ is located at $(V_w, \Phi_w + n_w\Phi_0)$, $n_w = j - 1$; those working points are defined as

$$V_w = f(I_b, \Phi_w + n_w\Phi_0); \Phi_w \in [0, \Phi_0) \tag{19}$$

This $n_w$ records the period where the working point is located, with which the flux-voltage function in (16) can be rewritten as

$$V_s = f_{V\text{-}\Phi}(\Phi_a - n_w\Phi_0); (\Phi_a - n_w\Phi_0) \in [0, \Phi_0) \tag{20}$$

The transfer function of this dc-SQUID is modeled as shown in Fig. 20(b), where the $n_w$ is automatically tranced by a virtual flux-quantum counter (FQC). The SQUID-based FLL is accordingly a feedback system model, as shown in. Fig. 20(c).

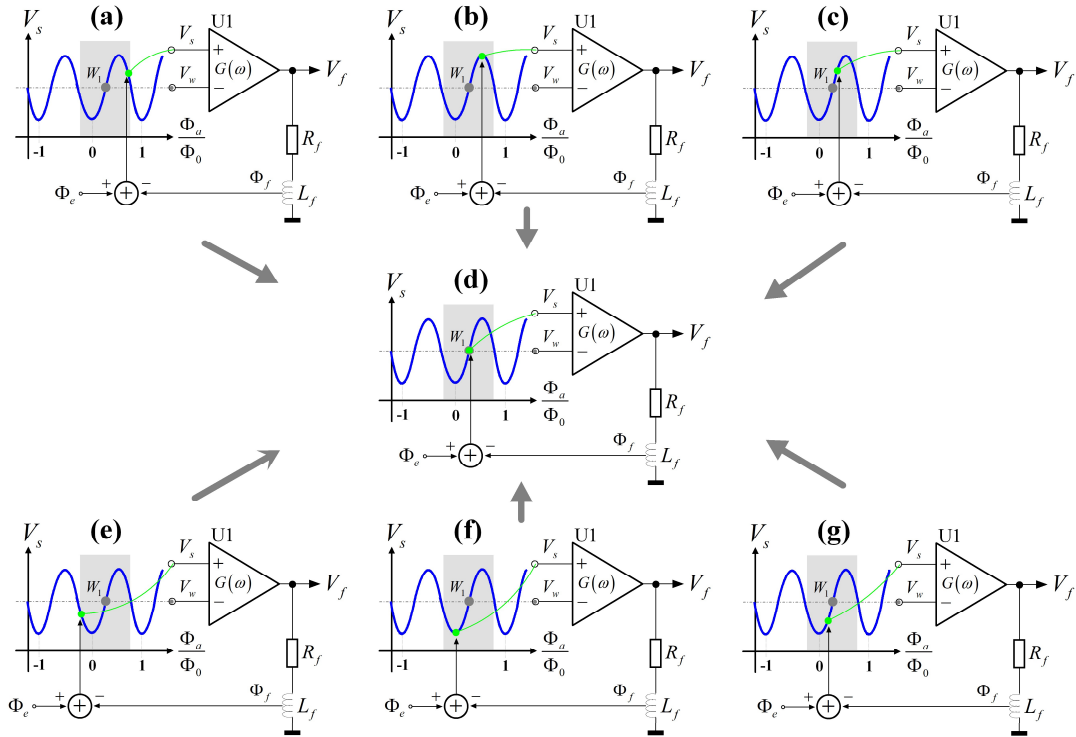

Fig. 21. Dynamic responses of dc-SQUID FLL; in the flux-voltage curve, $W_1$ $(V_w, \Phi_w)$ is the working point, and the point $(V_s, \Phi_a)$ is the current state of dc-SQUID: (a) the initial state $(V_s, \Phi_a)$ of dc-SQUID is at the negative slope, and $V_s > V_w$; (b) the initial state $(V_s, \Phi_a)$ is at the positive peak; (c) the initial state $(V_s, \Phi_a)$ is at the positive slope, and $V_s > V_w$; (d) the initial state $(V_s, \Phi_a)$ is coincide with $W_1$; (e) the initial state $(V_s, \Phi_a)$ is at the negative slope, and $V_s < V_w$; (f) the initial state $(V_s, \Phi_a)$ is at the negative peak; (g) the initial state $(V_s, \Phi_a)$ is at the positive slope, and $V_s < V_w$.

The responses of FLL to different initial states of $\Phi_a$ are illustrated in Fig. 21, where the dc-SQUID is depicted by its flux-voltage curve in $\Phi_a$ and $V_s$ coordinates. The increase of $\Phi_e$ will push $\Phi_a$ to right, while the increase of $\Phi_f$ will push $\Phi_a$ to left; the output $V_s$ is amplified by the non-inverting input of U1 with gain $G(\omega)$; it will increase $V_f$ and $\Phi_f$, if $V_s > V_w$, while decrease $V_f$ and $\Phi_f$, if $V_s < V_w$. The working principle of this system is simply described as followings:
1) If $V_s > V_w$, the room-temperature amplifier will increase $V_f$ and $\Phi_f$ to push $\Phi_a$

back to left, as shown in Fig. 21(a), (b) and (c).
2) If $V_s < V_w$, the room-temperature amplifier will decrease $V_f$ and $\Phi_f$ to pull $\Phi_a$ back to right; as shown in Fig. 21(e), (f) and (g).
3) If $V_s = V_w$, the room-temperature amplifier will keep $\Phi_a$ stable at $\Phi_w$, as illustrated in Fig. 21(d).

As long as the initial state ($V_s$, $\Phi_a$) is within the gray area around $W_1$ shown in Fig. 21, the FLL will always move the $\Phi_a$ back to $\Phi_w$. In other words, the FLL will be relocked at $\Phi_w + n_w\Phi_0$, when the initial $\Phi_a$ is in the grey area around $\Phi_w + n_w\Phi_0$.

In Fig. 21, the various initial states of $\Phi_a$ are set by the sudden change of $\Phi_e$; the relocking of FLL exhibits the step-response of FLL to the step input of $\Phi_e$. The response time of FLL depends on the deflection of $\Phi_a$:
1) The response time is approximately linear to the deviation of $\Phi_a$, if the state ($V_s$, $\Phi_a$) is located at the positive slope, as shown in Fig. 21(b), (c), (f) and (g).
2) The response time is nonlinear to the deviation of $\Phi_a$, if the state ($V_s$, $\Phi_a$) is located at the negative slope, as shown in Fig. 21(a) and (e).

In summary, a SQUID-based FLL is a flux-follower implemented by a SQUID-based FF-OPA. In the FF-OPA, the dc-SQUID is an AFE with a periodical nonlinear flux-to-voltage characteristic to sense flux inputs. The SQUID-based flux-follower is similar with the semiconductor FET-based voltage followers, except that the SQUID-based analog-front-end (AFE) has multiple working points in its flux-to-voltage characteristic while the FET-based AFE usually has only one working point.

### 6.3 Transfer function of FLL

In the flux-voltage characteristic shown in Fig. 22 (a), the transfer function of dc-SQUID in response to the small signal input around $W_1$, is approximately defined with the transfer coefficient $\partial V_s/\partial \Phi_a$ at $W_1$; the $\partial V_s/\partial \Phi_a$ at working points is defined by the flux-voltage function as

$$\frac{\partial V_s}{\partial \Phi_a} = \frac{\partial f(i_b, \Phi_a)}{\partial \Phi_a}\bigg|_{\substack{i_b=I_b \\ \Phi_a=\Phi_w+n_w\Phi_0}} \quad (21)$$

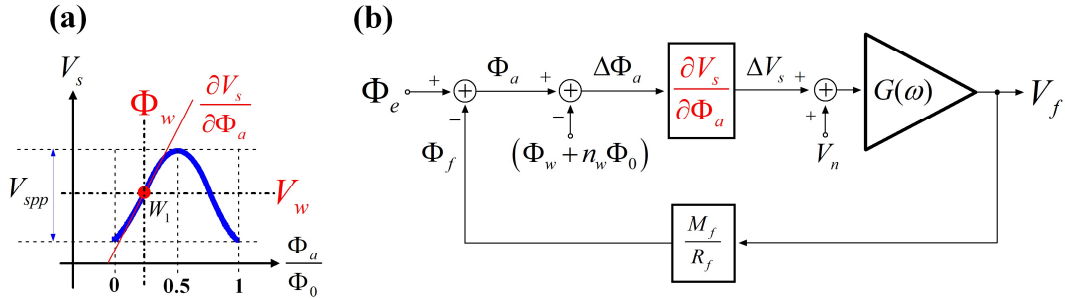

Fig. 22. (a) Transfer coefficient at $W_1$; (b) system diagram of FLL locked at $W_1$ [26].

With this small signal transfer function, the linear transfer function diagram of the FLL is shown in Fig. 22(b). In this model, the $\Phi_a$ is varying around $\Phi_w$ with a deflection

$\Delta\Phi_a$; the dc-SQUID outputs a $\Delta V_s$ like a linear flux-to-voltage convertor, as long as the $|\Delta\Phi_a|$ is smaller than a value $\varepsilon$ ($\varepsilon < 0.5\Phi_0$), $|\Delta\Phi_a| < \varepsilon$.

The negative feedback mechanism of the FLL keeps canceling the $\Delta V_s$, and $\Phi_a$ will be fixed at the working point dynamically, thus

$$\left.\begin{array}{l} \Delta V_s + V_n \approx 0 \\ \Delta V_s = \dfrac{\partial V_s}{\partial \Phi_a}\Delta\Phi_a \\ \Delta\Phi_a = \Phi_a - \Phi_w - n_w\Phi_0 \end{array}\right\} \Rightarrow \Phi_a = \Phi_w + n_w\Phi_0 - \Phi_n \quad (22)$$

where the $\Phi_n$ is the equivalent flux noised contributed by the voltage noise $V_n$ of the room-temperature amplifier, and

$$\Phi_n = \frac{V_n}{\partial V_s / \partial \Phi_a} \quad (23)$$

Accordingly, the flux-voltage conversion function of FLL in static state is

$$\Phi_e = \Phi_f + \Phi_a = \Phi_f + \Phi_w + n_w\Phi_0 - \Phi_n \quad (24)$$

Meanwhile, to a given $V_{f\_FS}$ which is the full-scale range of $V_f$, the full-scale range of the input flux, namely $\Phi_{e\_FS}$ will be

$$\Phi_{e\_FS} = \frac{M_f}{R_f} V_{f\_FS} \quad (25)$$

For a limited $V_{f\_FS}$ in the measurement instruments and a fixed $M_f$ in the given SQUID chip, the $\Phi_{e\_FS}$ is only adjusted by $R_f$ in the practical operations.

In frequency domain, the transfer function of the FLL is defined by

$$H(\omega) = \frac{\Phi_f(\omega)}{\Phi_e(\omega)} = \frac{1}{1 + R_f / (R_{ftr} G(\omega))}; \omega > 0 \quad (26)$$

where $R_{ftr}$ is the trans-impedance of the dc-SQUID seen from the two terminals of the feedback coil, and

$$R_{ftr} = \frac{\partial V_s}{\partial \Phi_a} M_f \quad (27)$$

This $R_{ftr}$ is decided by the transfer coefficient of the SQUID.

From the transfer function, the cut-off frequency $f_c$ of the FLL can be defined as

$$\left| R_f / (R_{ftr} G(\omega)) \right| = 1 \Rightarrow f_c \approx \frac{R_{ftr}}{R_f} GBP(f_c) \quad (28)$$

where $GBP(f)$ is the gain-bandwidth-product (GBP) of the room-temperature amplifier with a gain $G(\omega)$, and

$$GBP(f) = |G(\omega)f| \quad (29)$$

The maximum slew-rate, $SR_{max}(f)$ [26], evaluates the response of the FLL to large input of $\Phi_e$; the maximum swing of $\Phi_f$ driven by the room-temperature amplifier is less than $V_{spp}G(\omega)M_f/R_f$, because the maximum swing of $V_s$ is limited by the $V_{spp}$. Accordingly, the maximum slew-rate of the $\Phi_f$ is limited by

$$SR_{\max}(f) \approx \pi V_{Spp} \frac{M_f}{R_f} GBP(f) \tag{30}$$

The $GBP(f)$ is a constant for the room-temperature amplifier implemented by one operational amplifier or one integrator; the maximum slew rate for the FLL with a constant $GBP(f)$ is decided by the cut-off frequency $f_c$ as [26]

$$SR_{\max}(f) \approx \frac{\pi V_{Spp} f_c}{\partial V_s/\partial \Phi_a} \approx \frac{\pi \Phi_0 f_c}{2} \tag{31}$$

However, the $f_c$ is limited by the phase-margin (PM) of the $G(\omega)$, and the PM is reduced rapidly in the cascaded amplifier; this is why we suggest to design a FLL with no more two amplifiers [24].

### 6.4 How to improve performances of FLL

The advantages of SQUID sensors are low noise, low drift, wide range, and high slew-rate; they are achieved by using the different readout schemes as summarized in Table 2. Those schemes can be interpreted by the transfer functions of FLL.

Table 2. Schemes to improve the performance of FLLs

| Performance | Scheme | References |
|---|---|---|
| Low noise | 1) To lower $V_n$ of preamplifier | [46-51] |
|  | 2) To improve $\partial V_s/\partial \Phi_a$ | [14],[52],[53], [54-58] |
|  | 3) Using intermidiate amplifier | [25],[59-63] |
| Low drift | 1) To reduce $1/f$ noise in $\Phi_n$ | [25], [64],[65] |
|  | 2) Supressing the drift of $\Phi_w$ | [66-68] |
| Wide range | To trace the variation of $n_w$ | [69-74] |
| High slew-rate | To improve $GBP(f)$ | [24], [75-79] |

# 7. Bias modes and working points

### 7.1 Bias and amplifier circuits

For a given dc-SQUID, the transfer coefficient depends on the location of working point in the current-voltage curves; the working point is set by room-temperature amplifier. Two schemes of the room-temperature amplifier in non-inverting and inverting modes are exhibited in Fig. 23, where dc-SQUID is read out by the preamplifier OPA1 with $R_g$ connected between the output and the inverting input; the OPA2 is working with an open-loop gain; the GBP is modified by $R_2$ and $C_2$.

The dc-SQUID in Fig. 23(a) is working with the preamplifier in the current-bias-voltage amplifier (VBCA) mode [80], as shown in Fig. 24(a); it shown in Fig. 23(b) is working in the voltage-bias-current-amplifier (VBCA) mode [80], as shown in Fig.

24(b). The $R_{wire}$ is the resistance of the wire connected between the dc-SQUID and the preamplifier; the $R_g$ is much larger than $R_r$, $R_{wire}$, as well as the shunt resistors of the dc-SQUID.

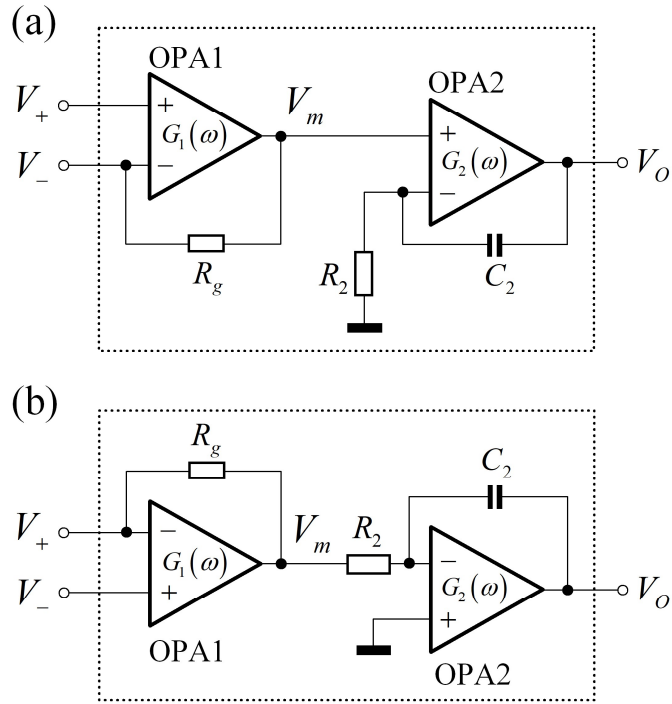

Fig. 23. Two room-temperature amplifier schemes: (a) noninverting mode, and (b) inverting mode.

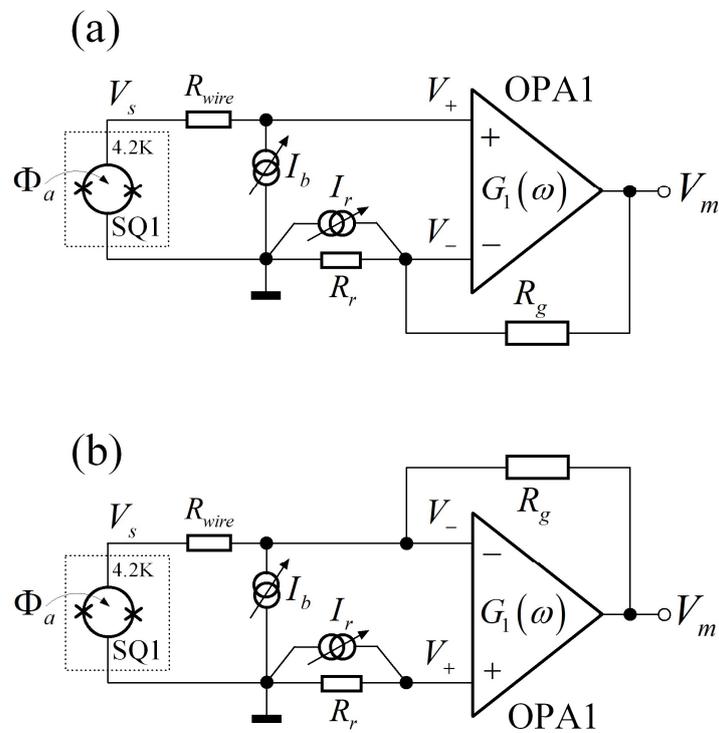

Fig. 24. Two bias-and-amplifier modes: (a) CBVA mode, and (b) VBCA mode [80].

In the CBVA mode, the $V_s$ of the dc-SQUID under the constant bias current $I_b$ is linearly amplified to $V_m$; the transfer function from $V_s$ to $V_m$ is

$$\left(V_m/R_g + I_r\right)R_r = V_s + R_{wire}I_b \tag{32}$$

In the VBCA mode, the $i_b$ flowing through the dc-SQUID under the bias voltage $I_rR_r$ is linearly amplified to $V_m$; the transfer function from $i_b$ to $V_m$ is

$$V_m/R_g + I_b = i_b \tag{33}$$

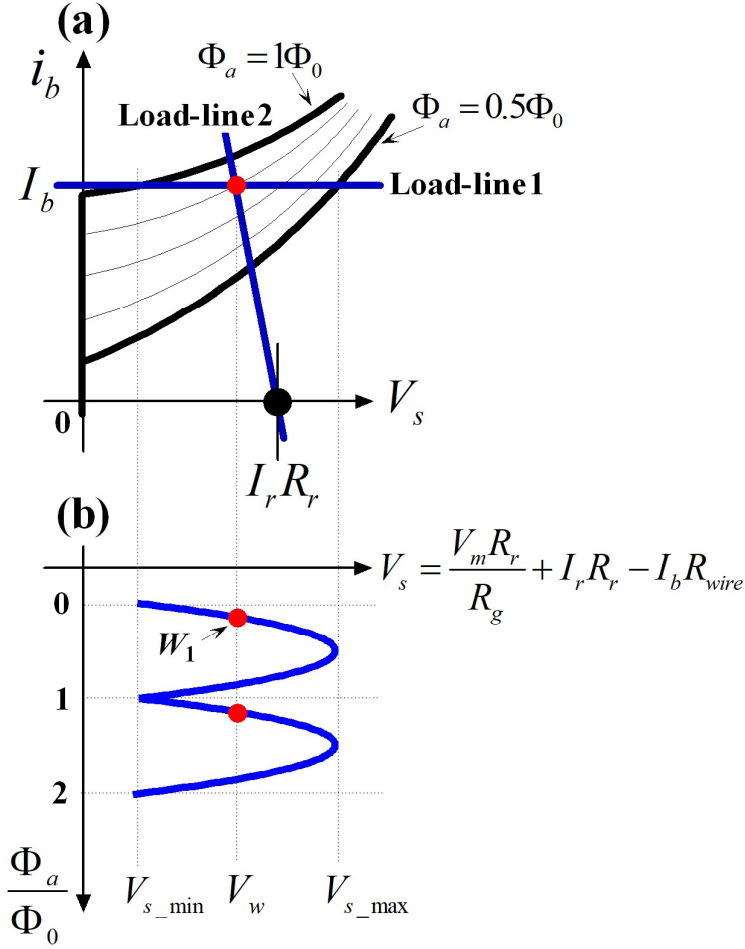

Fig. 25. Principle of CBVA mode: (a) two load lines and (b) flux-voltage characteristic [81].

### 7.2 Working points set by two bias modes

When the FLL based on either the CBVA or the VBCA mode is in the locking state, the output of OPA1 will be kept at zero, $V_m = 0$; the current flowing into $R_g$ is ignored, and the $i_b$ is totally supplied by the current source $I_b$; the $V_+$ and $V_-$ at two inputs of the OPA1 are equal, $V_+ = V_-$. Therefore, the working point of the dc-SQUID in either bias mode is selected by two load lines:

The first load line, which is named as Load-line1 [81], is

$$i_b = I_b \tag{34}$$

The second load lines, which is called Load-line2 [81], is

$$V_s + i_b R_{wire} = I_r R_r \qquad (35)$$

Two load lines set in CBVA mode together with the current-voltage characteristics of the dc-SQUID are drawn in Fig. 25(a). The Load-line1 is moved up and down by adjusting $I_b$, while the Load-line2 is shifted from left to right by increasing $I_r$.

When the FLL is in opened-loop state, the flux-voltage characteristic of the dc-SQUID is read out by the OPA1 in CBVA mode, which is the projection of the current-voltage characteristics on the Load-line1, as illustrated in Fig. 25(b), where the intersection of two load lines is accordingly projected to the working points.

Similarly, two load lines set in VBCA mode together with the current-voltage characteristics of the dc-SQUID are drawn in Fig. 26(a). When the FLL is in opened-loop state, the OPA1 in VBCA mode reads out the flux-current characteristic of the dc-SQUID as illustrated in Fig. 26(b); it is the projection of the current-voltage characteristics on the Load-line2, and the working points is set by the intersection of two load lines.

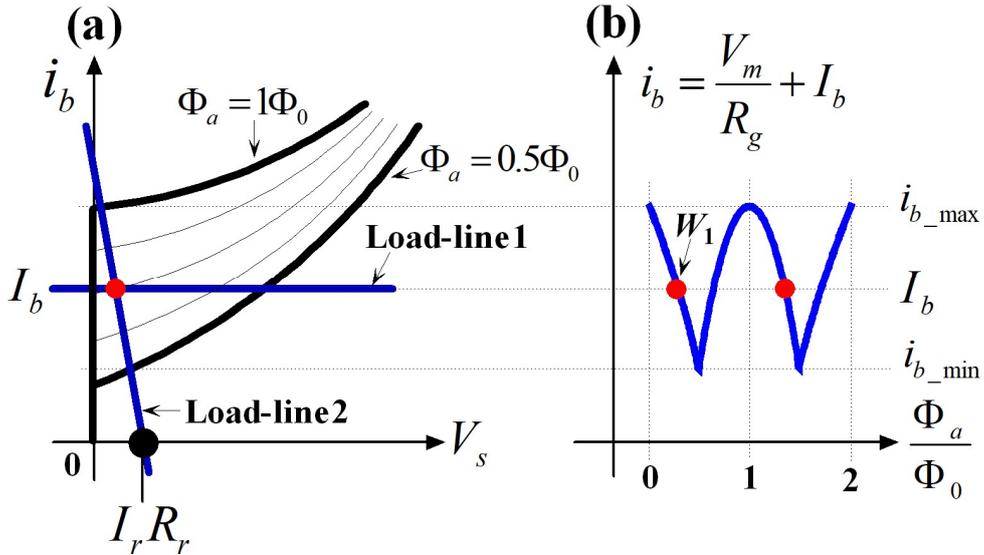

Fig. 26. Principle of VBCA mode: (a) two load lines and (b) flux-current characteristic [81].

In operations of FLL, the working point of dc-SQUID is simply set with two practical principles:
1) The first one is to achieve the maximum swing of $V_m$ scanned by the input $\Phi_a$; the swing of $V_m$ is adjusted by the Load-line1 in CBVA mode; it is adjusted through Load-line2 in VBCA mode.
2) The second one is to set the zero-baseline of $V_m$ in the middle of the swing. The baseline of of $V_m$ is set by the Load-line2 in CBVA mode, and is adjusted by Load-line1 in VBCA mode.

By comparing Fig. 25 with Fig. 26, we can find that the working points selected by two bias modes will be in different locations in the current-voltage curves.

## 7.3 Small signal analyses

In either CBVA or VBCA circuit, the small-signal response of the dc-SQUID around the selected working point is described as

$$\Delta V_s = f(I_b + \Delta i_b, \Phi_w + \Delta \Phi_a) - f(I_b, \Phi_w) \approx R_d \Delta i_b + \frac{\partial V_s}{\partial \Phi_a} \Delta \Phi_a \quad (36)$$

where $R_d$ is the dynamic resistance of the dc-SQUID working at $W_1$; it is defined by

$$R_d = \left. \frac{\partial f(i_b, \Phi_a)}{\partial i_b} \right|_{\substack{i_b = I_b \\ \Phi_a = \Phi_w + n_w \Phi_0}} \quad (37)$$

The Thevenin's equivalent circuit of the dc-SQUID in small signal mode, is equivalent to a flux-controlled voltage source in serial with an internal resistor $R_d$, as shown in Fig. 27. The voltage signal generated by $\Delta \Phi_a$ is amplified by a non-inverting amplifier in the CBVA scheme, as shown in Fig. 27(a), and it is amplified by an inverting amplifier in the VBCA scheme, as shown in Fig. 27(b).

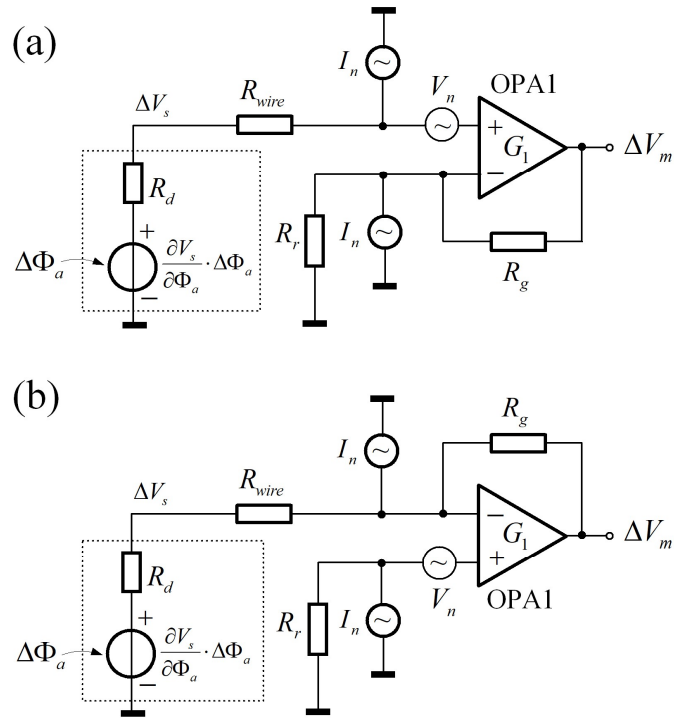

Fig. 27. Small signal circuit of dc-SQUID working (a) in CBVA mode, and (b) in VBCA mode [80].

In both CBVA and VBCA circuits, the current noise $I_n$ and voltage noise $V_n$ of the room-temperature preamplifier OPA1 will contribute an equivalent flux noise $\Phi_n$ to the FLL as

$$\Phi_n = \frac{\sqrt{V_n^2 + (I_n R_d + I_n R_{wire})^2 + (I_n R_r)^2}}{\partial V_s / \partial \Phi_a} \quad (38)$$

The voltage noise $V_n$ of room-temperature amplifier is the dominant noise source, if

the $R_r$, $R_d$, and $R_{wire}$ meet the impedance matching condition [80], namely

$$R_r < R_n; (R_d + R_{wire}) < R_n; R_n = V_n/I_n \qquad (39)$$

where $R_n$ is the so-called noise-impedance of the preamplifier.

Therefore, in both bias and amplifier modes, the $\Phi_n$ contributed by the $V_n$ is only decided by the $\partial V_s/\partial \Phi_a$ at the selected working point.

### 7.4 Comparison between two bias modes: an example

In practical operations, the two bias modes will set their working points at different locations in the current-voltage curves, and will therefore achieve different noise performances, if different working points achieve different $\partial V_s/\partial \Phi_a$. For a practical dc-SQUID, we can measure its current-voltage curves as shown in Fig. 28(a); the contour map shown in Fig. 28(b) exhibits the value of $\partial V_s/\partial \Phi_a$ at each point in the current-voltage curves; the higher value of $\partial V_s/\partial \Phi_a$ is highlighted with darker color [81].

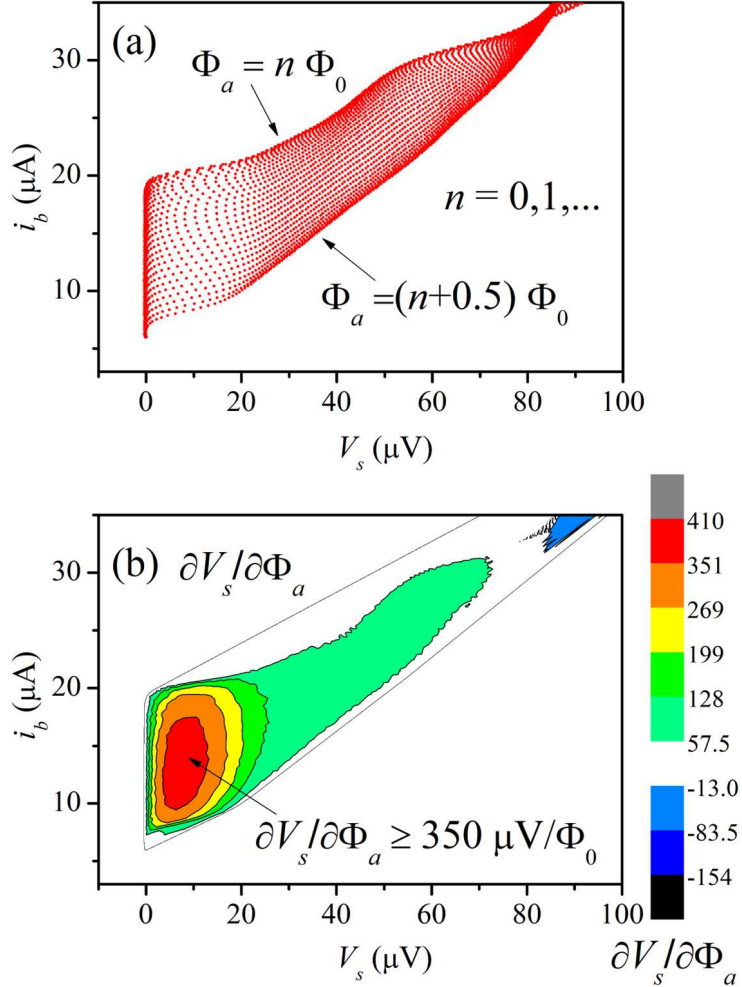

Fig. 28. (a) Current-voltage characteristics measured from a practical dc-SQUID; (b) The contour map of flux-to-voltage transfer coefficient [81].

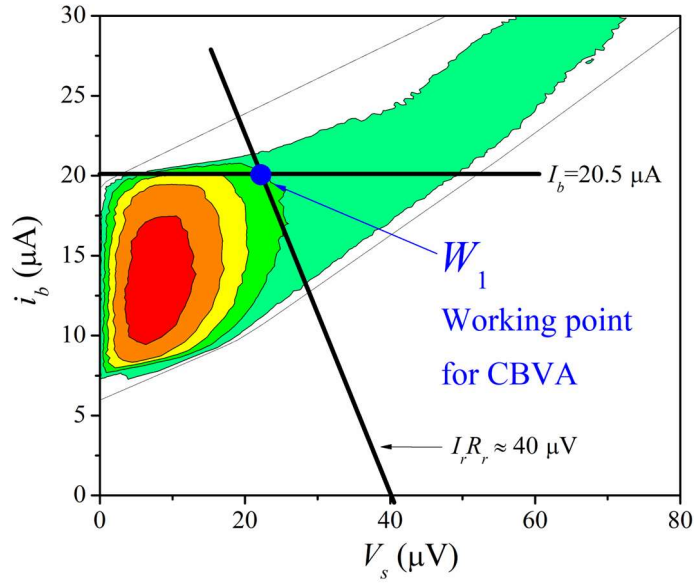

Fig. 29. Working point selected by CBVA mode [81].

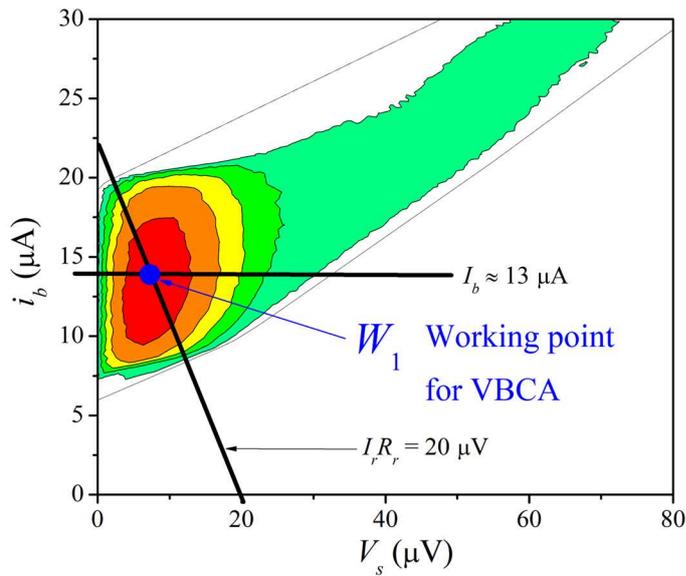

Fig. 30. Working point selected by VBCA mode [81].

Fig. 29 depicts the working point selected by the CBVA mode in the contour map, where the working point at the intersection point of two load lines is away from the area with high transfer coefficient. In the contrast, the working point selected by the VBCA mode is located in the area with high transfer coefficient, as shown Fig. 30.

The noise spectra of the FLLs in two bias schemes are compared as shown in Fig. 31. The FLL with VBCA scheme achieves better noise performance than the one with CBVA scheme, because the working point selected by the VBCA scheme achieve higher transfer coefficient than the one selected by the CBVA scheme. More details are seen in the reference [81].

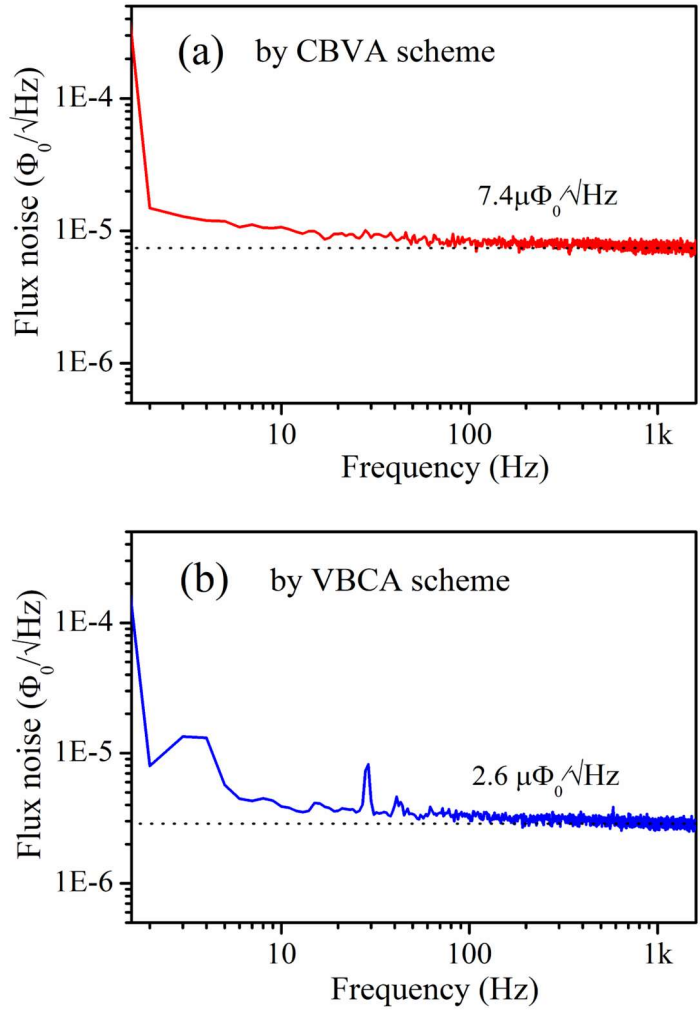

Fig. 31. (a) Noise spectrum read out by CBVA mode; (b) noise spectrum read out by VBCA mode [81].

# 8. Improving flux-voltage transfer coefficient

## 8.1 Four low-noise readout schemes

To further improve the transfer coefficient of dc-SQUID in the CBVA and VBCA circuits, the near-SQUID feedback schemes, such as additional positive feedback (APF) scheme [54], noise cancellation scheme [56], bias-current feedback (BCF) scheme [57], SQUID bootstrap circuit (SBC) scheme [58], are developed, as illustrated in Fig. 32. Four readout schemes are summarized in Table 3.

Table 3. Four direct readout schemes with near-SQUID feedback

|  | Near-SQUID feedback | Bias and amplifier mode |
| --- | --- | --- |
| APF sheme | Voltage feedback | CBVA |
| NC scheme | Voltage feedback | VBCA |
| APF and BCF scheme | Volage and current feedback | CBVA |
| SBC scheme | Volage and current feedback | VBCA |

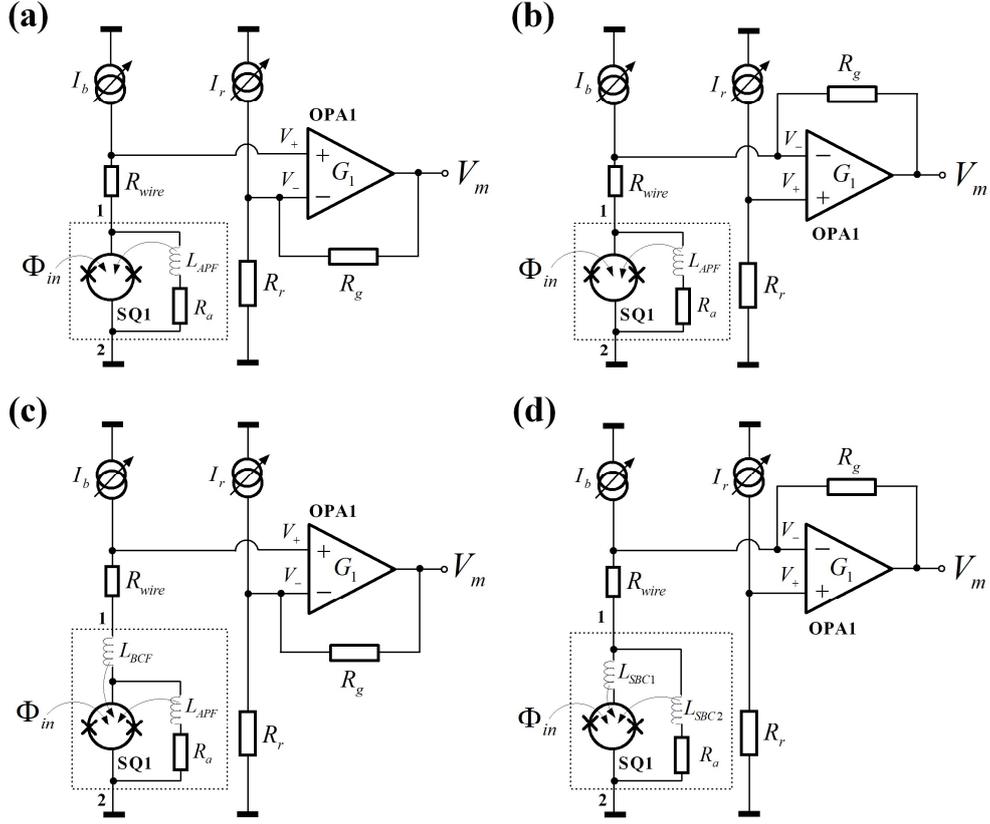

Fig. 32. Four readout schemes with near-SQUID feedback: (a) APF scheme working in CBVA mode []; (b) APF scheme working in VBCA mode []; (c) APF combined with BCF scheme working in CBVA mode []; (d) SBC scheme working in VBCA mode [80].

## 8.2 General equivalent circuit of near-SQUID feedback

Three kinds of near-SQUID feedback schemes in Fig. 32 are extracted and depicted in Fig. 33. Their common equivalent circuit is shown in Fig. 33(d), where the current $i_b$ is turned into the flux $M_I i_b$ in the dc-SQUID through a mutual inductance $M_I$, and the voltage $V_s$ generates the flux $M_V V_s/R_a$ to the dc-SQUID through a mutual inductance $M_V$. The $M_I$ and $M_V$ in different schemes are configured with different values as

$$\begin{bmatrix} M_I \\ M_V \end{bmatrix} = \begin{cases} \begin{bmatrix} 0 \\ M_{APF} \end{bmatrix} ; \text{in APF scheme} \\ \begin{bmatrix} M_{BCF} \\ M_{APF} \end{bmatrix} ; \text{in APF and BCF scheme} \\ \begin{bmatrix} M_{SBC1} \\ M_{SBC1} + M_{SBC2} \end{bmatrix} ; \text{in SBC scheme} \end{cases} \qquad (40)$$

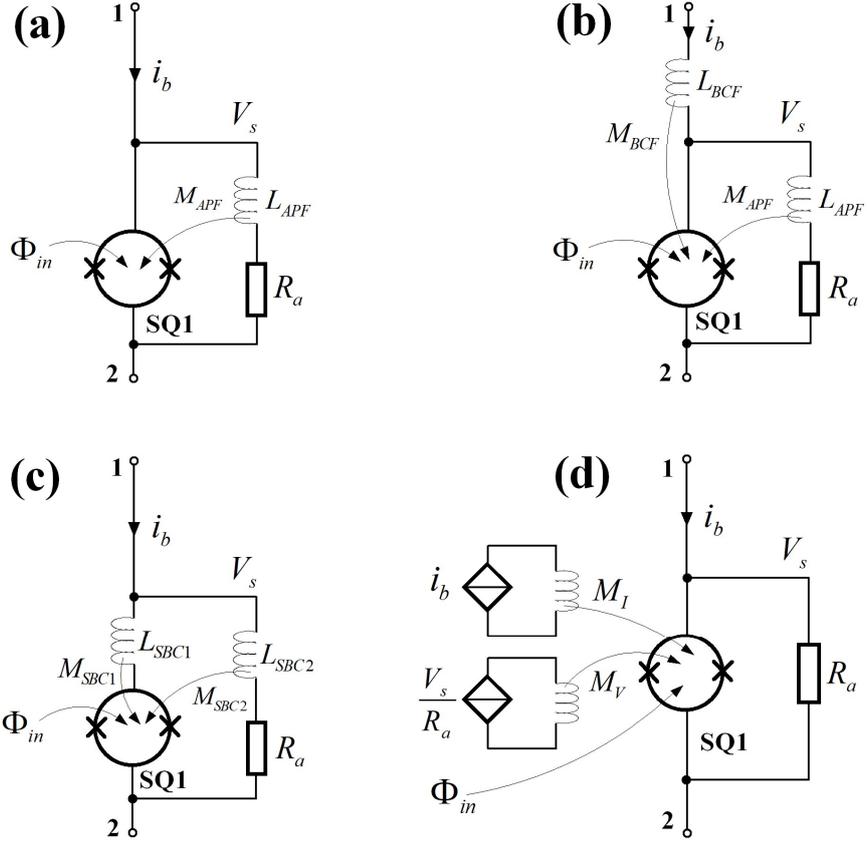

Fig. 33. Near-SQUID feedback schemes and general equivalent circuit: (a) APF scheme; (b) APF and BCF scheme; (c) SBC scheme; (d) general equivalent circuit.

*8.3 Small-signal model of near-SQUID feedback circuits*

In the locked FLL, the near-SQUID feedback schemes are also working in small signal mode, which small-signal circuit is shown in Fig. 34(a). The feedback effects of $\Delta i_b$ and $\Delta V_s$ are equivalent to a current-controlled voltage source and a voltage-controlled voltage sources, as shown in Fig. 34(b), in which, $R_I$ and $R_V$ are two equivalent trans impedances [80] defined as

$$\begin{cases} R_I = \dfrac{\partial V_s}{\partial \Phi_a} M_I \\ R_V = \dfrac{\partial V_s}{\partial \Phi_a} M_V \end{cases} \quad (41)$$

The final Thevenin's equivalent circuit of the near-SQUID feedback schemes in small-signal mode is shown in Fig. 34(c). The overall transfer coefficient and the dynamic resistance are derived as follows [80]:

$$\frac{\partial V_s}{\partial \Phi_{in}} = \frac{\partial V_s}{\partial \Phi_a} \cdot \frac{1}{1-(R_V - R_d)/R_a} \quad (42)$$

$$(R_d)^* = R_d \cdot \frac{1 - R_I/R_d}{1-(R_V - R_d)/R_a} \quad (43)$$

where the stable condition for the feedback schemes is

$$R_I < R_d; R_V < R_d + R_a \qquad (44)$$

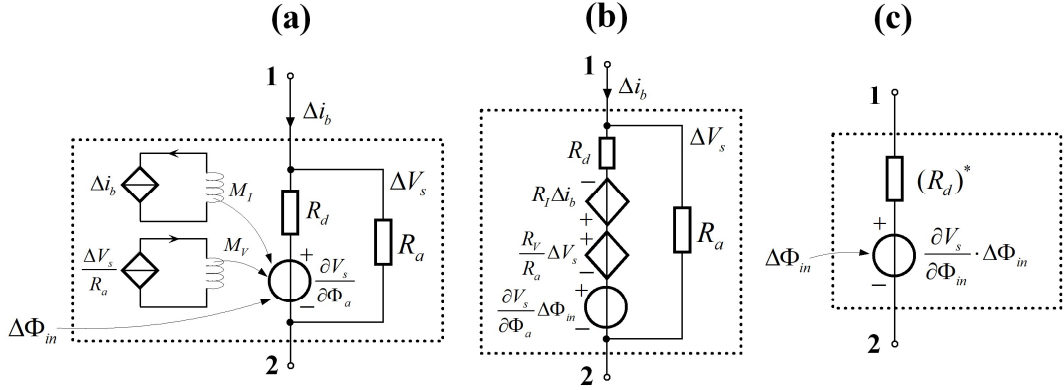

Fig. 34. (a) small-signal model of near-SQUID feedback scheme; (b) simplified equivalent circuit based on TIA model; (c) The final Thevenin's equivalent circuit [].

We find that the near-SQUID schemes modify the transfer coefficient and dynamic resistance through two trans impedances $R_I$ and $R_V$:
1) The $R_V$ is the only parameter that can improve the transfer coefficient. The increase of $R_V$ will accordingly increase both the transfer coefficient and the dynamic resistance.
2) The $R_I$ has no effect on the transfer coefficient; it is only used to adjust the dynamic resistance for the noise matching condition in (39).

In summary, three near-SQUID feedback schemes are equivalent; the APF branch has the effect to increases the overall transfer coefficient, whereas the coil $L_{BCF}$ and the coil $L_{SBC1}$ are only useful in modifying the dynamic resistance.

## 9. Working point jump and flux-quanta counting

### 9.1 Working point shift in FLL

The working point shift in a FLL is illustrated in Fig. 35. In the flux-voltage curve of dc-SQUID, there is a threshold value $\Phi_{th}$ between $\Phi_w$ and $1\Phi_0$, where $V_s > V_w$, if $\Phi_w < \Phi_a < \Phi_{th}$, and $V_s < V_w$, if $\Phi_{th} < \Phi_a < 1\Phi_0$. If a sudden change of $\Phi_e$ move $\Phi_a$ to the region between $\Phi_{th}$ and $1\Phi_0$ (colored in yellow), as shown in Fig. 35 (a), the FLL will pull the $\Phi_a$ to $W_2$, as exhibited in Fig. 35(b).

After the working point shifts from $W_1$ to $W_2$, $n_w$ is changed from 0 to 1, and the FLL will have a flux jump of $1\Phi_0$ in $\Phi_f$, according to (24). A typical flux jump measured in experiments is illustrated in Fig. 36, where the jump of $1\Phi_0$ in $\Phi_f$ is resulted by one flux-quantum increasement in $n_w$.

Therefore, to extend the range of FLL, we can track the change of $n_w$, $\Delta n_w$, from the flux jumps of $\Phi_f$, and include the $\Delta n_w$ in (24) to calculate the input $\Phi_e$. This read out method is called the flux-quantum counting (FQC) scheme [69-74].

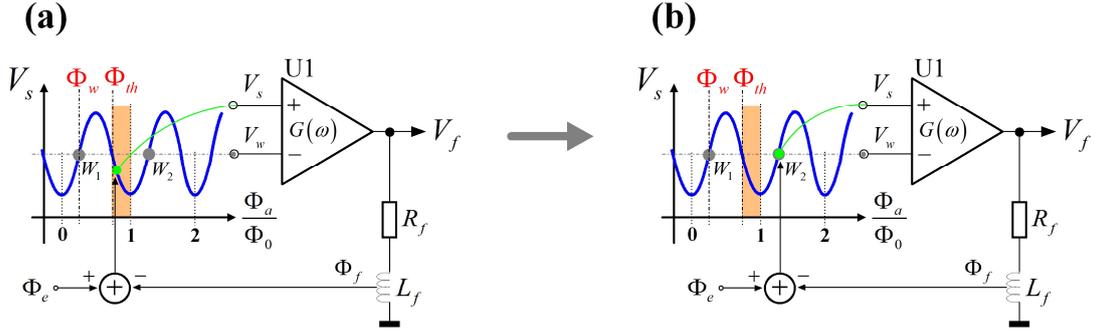

Fig. 35. Working point shift in FLL: (a) $\Phi_a$ is moved to the yellow area between $\Phi_{th}$ and $\Phi_0$ by the sudden change of $\Phi_e$; (b) FLL is relocked at $W_2$ with a flux jump of $1\Phi_0$ in $\Phi_f$.

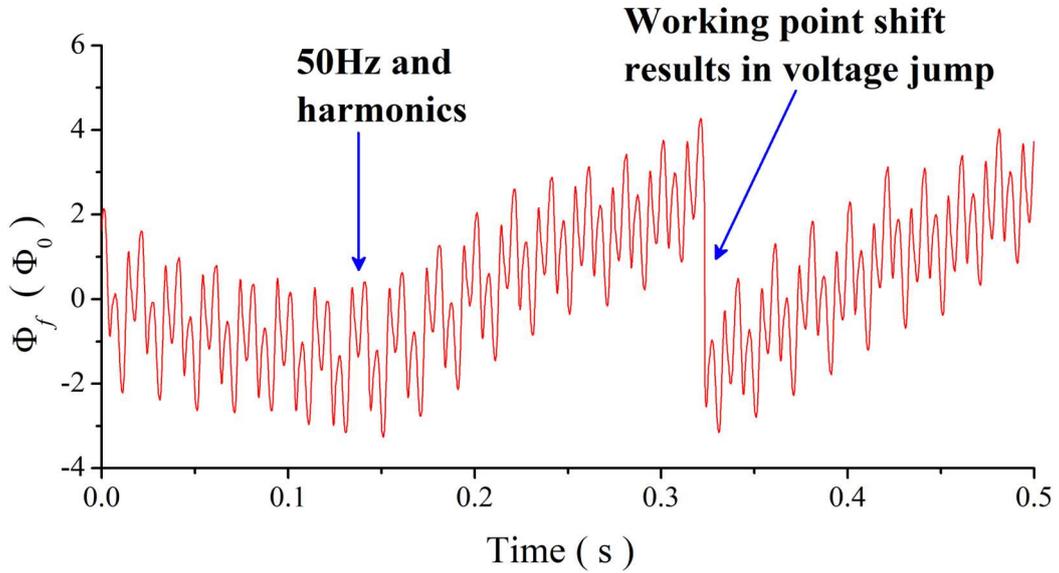

Fig. 36. A voltage-jump resulted by working point shift in a practical measurement.

### *9.2 Flux-quantum counting scheme*

The conventional FLL reads out the $\Phi_e$ with $\Phi_f$ by assuming that the $n_w$ in (24) remain unchanged, $\Delta n_w = 0$. If we reset the FLL intentionally when the output of FLL exceeds a threshold, and infer the $\Delta n_w$ according to the flux jumps in the measured FLL output, we can extend the measurement range of $\Phi_e$ by including both the results of $\Delta n_w$ and $\Phi_f$ in (24). This is exactly the principle that the conventional flux-quantum counting (FQC) readout schemes [69-71] are based to implement wide range measurements.

However, additional comparator and control circuits are used to reset the FLL; their response time is the dead time of FQC schemes, because the FQC scheme will miss counting the change of $n_w$ during the dead time. To reduce the dead time and avoid the miscounting of $n_w$, FQC schemes enhanced by simple proportional feedback are developed [73] [74]; their working point is reset automatically without any external control circuits.

### 9.3 Proportional feedback scheme

The proportional feedback schemes that reset the working point automatically are illustrated in Fig. 37. The one integrated with the dc-SQUID in one chip is demonstrated in Fig. 37(a), and the one implemented with a room-temperature proportional amplifier is illustrated in Fig. 37(c).

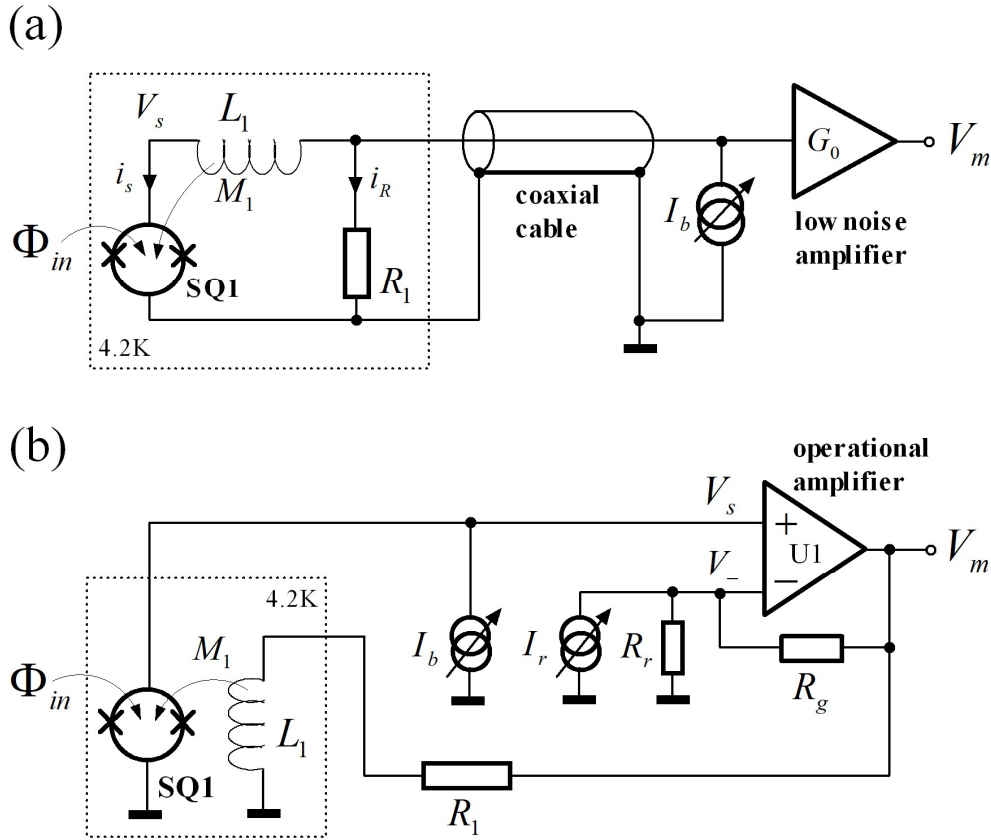

Fig. 37. Two proportional feedback schemes: (a) the one implemented with APF circuit in dc-SQUID chip [73]; (b) the one implemented with room-temperature preamplifier [74].

The proportional feedback schemes modify the flux-voltage characteristic of the dc-SQUID and narrow the $\Phi_{th}$ to $0.5\Phi_0$, as shown in Fig. 38. For a dc-SQUID with a flux-voltage curve shown in Fig. 38(a), the proportional feedback will pull the $\Phi_a$ in the region between $0.5\Phi_0$ and $1\Phi_0$ quickly back to the quasi-linear region between 0 and $0.5\Phi_0$; the region between $\Phi_{th}$ and $1\Phi_0$ is absent, and the $\Phi_{th}$ is narrowed to $0.5\Phi_0$.

If the scheme shown in Fig. 37(b) is used as the preamplifier circuit in a FLL, the proportional feedback scheme will keep the dc-SQUID working with only the quasi-linear characteristic between 0 and $0.5\Phi_0$, and will therefore reduce the response time of the FLL.

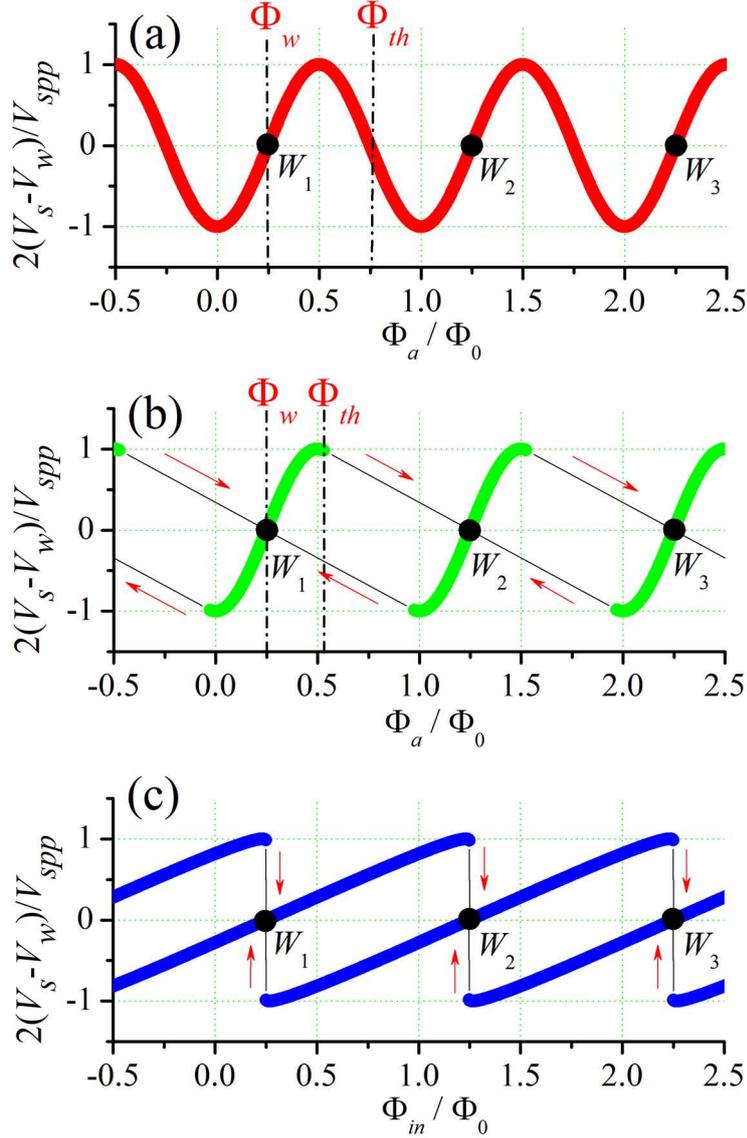

Fig. 38. (a) Original flux-voltage characteristic of dc-SQUID; (b) the flux-voltage characteristic of dc-SQUID modified by the proportional feedback scheme; (c) the overall flux-voltage characteristic read out by the proportional feedback scheme [73].

## 10. Other practical readout schemes

### 10.1 FLL with single operational amplifier

The room-temperature amplifier in the SQUID-based FF-OPA usually consists of a preamplifier and an integrator to achieve a high gain $G(\omega)$ [5], [20]. It is also implemented with a single amplifier [77] [78], as shown in Fig. 39, where the $R_g$ used to connect the inverting input and the output of U1 is switched by SW1 [24]. When SW1 is on, the FLL is same with the readout circuit shown in Fig. 37(b), and is used to set the working point according to its flux-voltage characteristic shown in Fig. 38(c). When SW1 is off, the circuit is in closed-loop mode, and is exactly the simple flux-follower shown in Fig. 19(b).

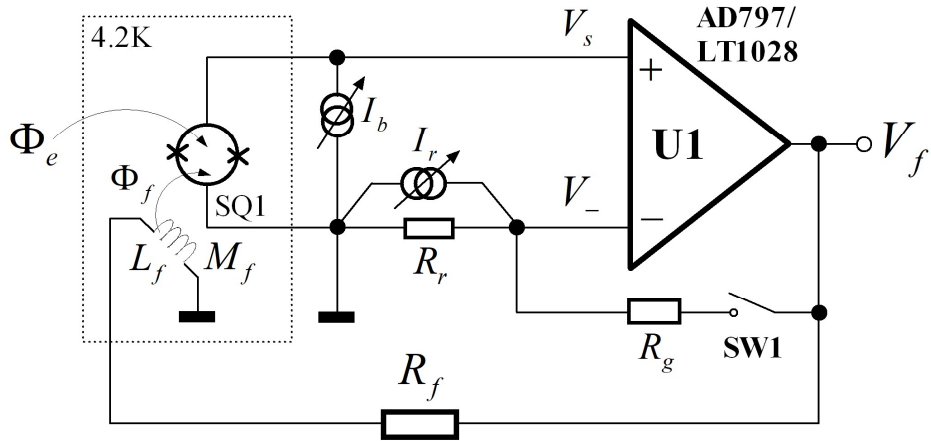

Fig. 39. Simple FLL using one operational amplifier [24].

## 10.2 Practical two-stage readout scheme

A practical two-stage readout scheme is exhibited in Fig. 40, where SQ1 is the first-stage dc-SQUID used to sense the flux input $\Phi_e$, and SQ2 is the second-stage dc-SQUID used to amplify the output of SQ1 [63].

Different with the conventional two-stage schemes [59-62], our two-stage scheme improves the transfer coefficient of two-stage SQUIDs with two feedback schemes:
1) The first one is the APF scheme nearby SQ1; it is used to improve the transfer coefficient of SQ1.
2) The second one the proportional feedback scheme; it is used to enable the second-stage SQUID amplifier to achieve quasi-linear flux-voltage characteristics shown in Fig. 38(c).

Two schemes are combined to achieve an improvement of 20dB in the overall flux-to-voltage transfer coefficient, and achieve a noise floor lower than $1\mu\Phi_0/\sqrt{Hz}$. More details are seen in the reference [63].

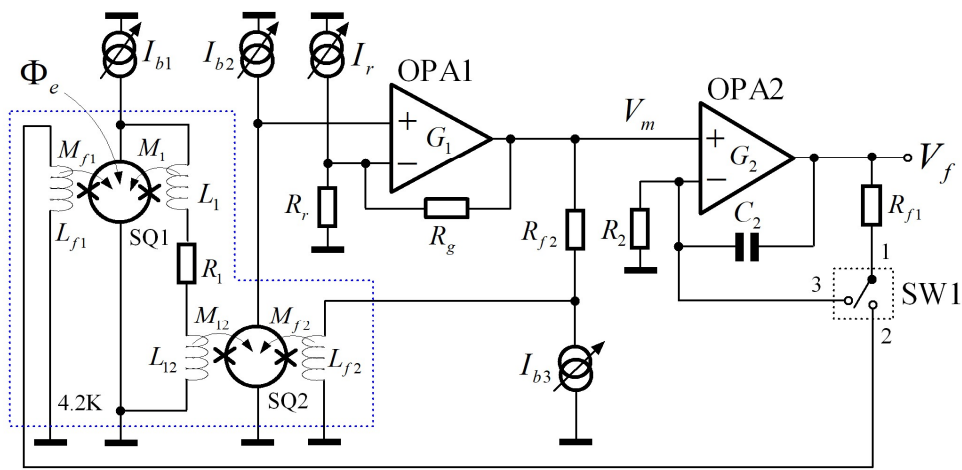

Fig. 40. FLL with two-stage SQUIDs, where SW1 is in 'off' state.

## 10.3 Suppression of the drift caused by the variation of $R_{wire}$

In the practical SQUID sensor systems that maintain cryogenic state with liquid Helium, the drop of the liquid Helium level will increase the $R_{wire}$ in the readout circuit shown in Fig. 41(a), and make the Load-line2 shown in Fig. 25 and Fig. 26 less steep. The effects induced by the increase of $R_{wire}$ are:

1) In the CBVA mode, increase of $R_{wire}$ will move the downward zero-baseline of $V_m$, and arouse the dc-offset drift in $\Phi_f$.
2) In the VBCA mode, increase of $R_{wire}$ will decrease the swing of $V_m$, and will also induce the dc-offset drift in $\Phi_f$.

A low differential amplifier scheme that suppresses the voltage offset created by $R_{wire}$ is shown in Fig. 41(b). Based on this scheme, we developed a low drift and compact practical SQUID-based MCG system, as seen in reference [68].

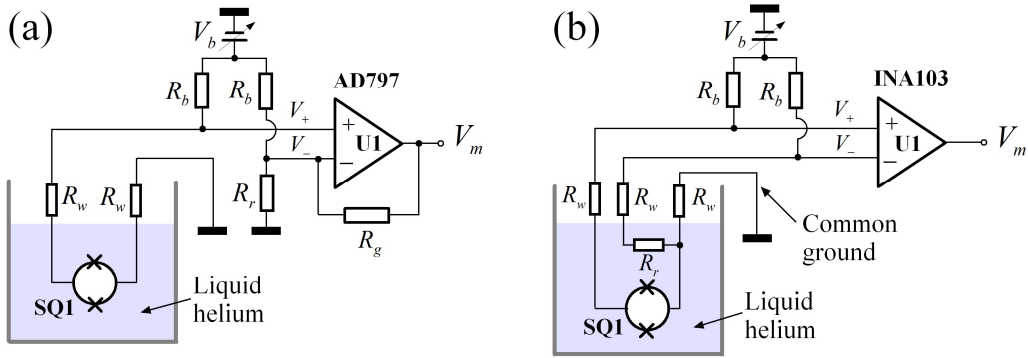

Fig. 41. Two direct readout schemes [68], where $R_w$ is the resistance of the single wire that connects the dc-SQUID to the room-temperature amplifier: (a) scheme using single-ended amplifier; (b) scheme using differential amplifier to cancel the voltages induced by $R_w$.

# 11. SQUID-based magnetic-field sensors

## 11.1 Concept of trans-impedance amplifier

A linear magnetic-field sensors is simply implemented by integrating a flux-transformer to a SQUID-based flux-follower, as demonstrated in Fig. 42. In Fig. 42(a), the flux-transformer transfer the flux signal coupled by its pick-up coil to the input of the flux-follower implemented by an FF-OPA; the flux-transformer and the feedback of the flux-follower can share the same input coil, as shown in Fig. 42(b).

Viewed at two terminals of the pick-up coil, the flux-flower integrated with a pick-up coil is exactly a trans-impedance-amplifier (TIA), with which, picked-up flux signal is linearly transformed into the voltage output $V_f$.

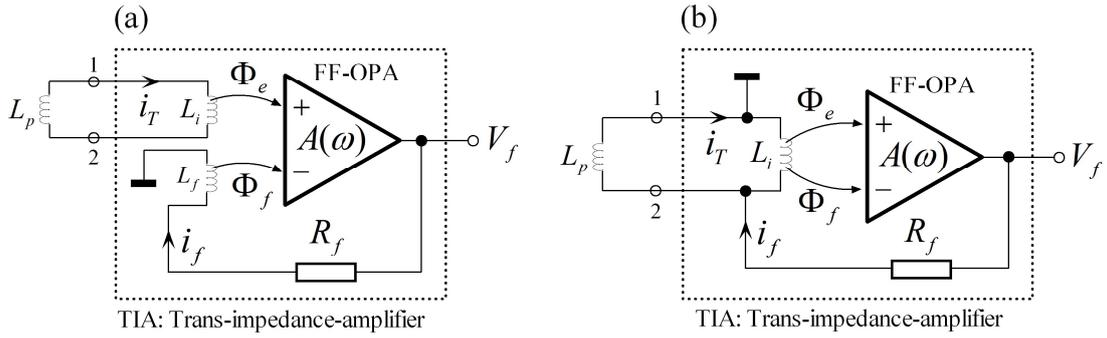

Fig. 42. Magnetic-field sensor implemented by FF-OPA and flux-transformer: (a) input flux and feedback flux are separately applied through two coils, (b) input flux and feedback flux are applied through one coil.

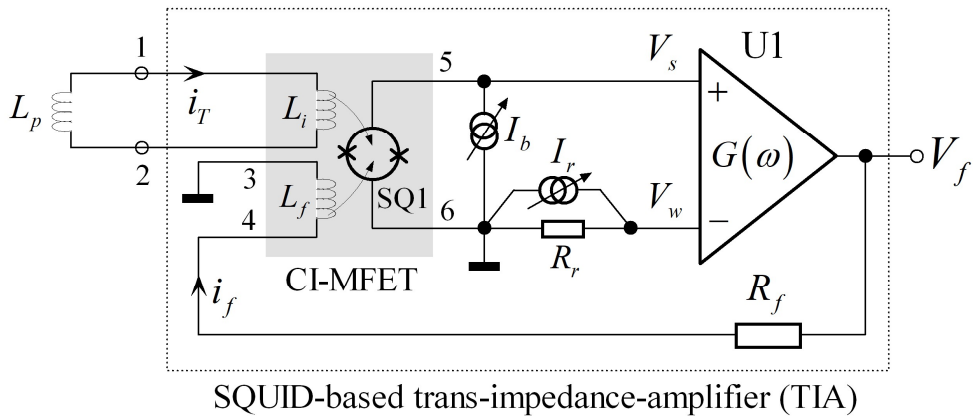

Fig. 43. SQUID-based TIA using a CI-MFET with two input coils.

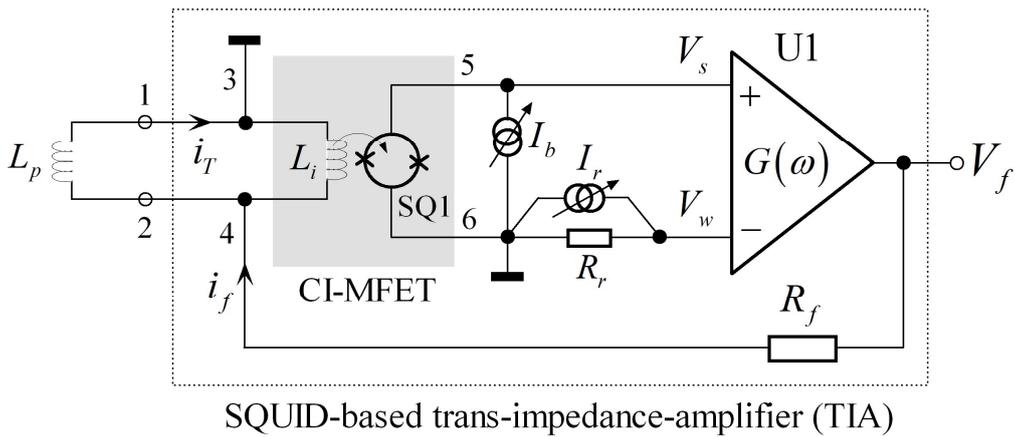

Fig. 44. SQUID-based TIA using a CI-MFET with one input coils.

*11.2 SQUID-based trans-impedance amplifier*

The scheme shown in Fig. 42(a) implemented with the practical SQUID-based FF-OPA is shown in Fig. 43, where, the dc-SQUID tightly coupled with two input coils is the current-input MFET (CI-MFET). The scheme shown in Fig. 42(b) is implemented with the SQUID-based FF-OPA, as shown in Fig. 44, where the CI-MFET has one current input.

The CI-MFETs have current inputs as the control gate, as the MOSFETs have the voltage inputs as the control gate. The input currents modulate the current-voltage characteristics of CI-MFET, as the voltage inputs modify the current-voltage characteristics of MOSFET.

*11.3 SQUID magnetic-field sensors*

SQUID magnetic-field sensors are simply assembled by connecting a pick-up coil to a SQUID-based TIA, as illustrated in Fig. 45. The style of pick-up coils decides what kind of magnetic fields the SQUID magnetic-field sensor is measuring.

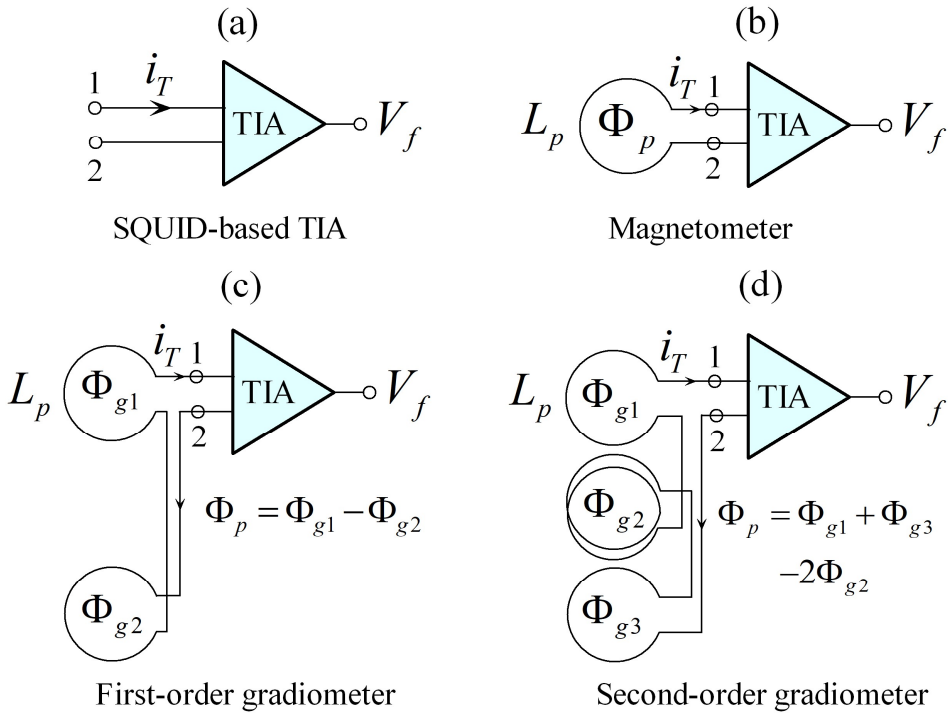

Fig. 45. SQUID sensors: (a) symbol of SQUID-based TIA; (b) magnetometer; (c) first-order gradiometer; (d) second-order gradiometer.

# 12. Multichannel SQUID system

*12.1 Example of Multi-channel SQUID system*

A Magneto-cardio-gram (MCG) system based on multi-channel SQUID sensors is exhibited in Fig 46. Exited by the heartbeats, the cardio neuro currents generate magnetic fluxes coupled by the pick-up coils. If they are modeled as a group of loop currents, $j_{x1}$, …, $j_{xQ}$, the MCG system is exactly a contactless multi-channel current measurement system, which senses the target currents through mutual inductances, and turns them linearly into voltage outputs.

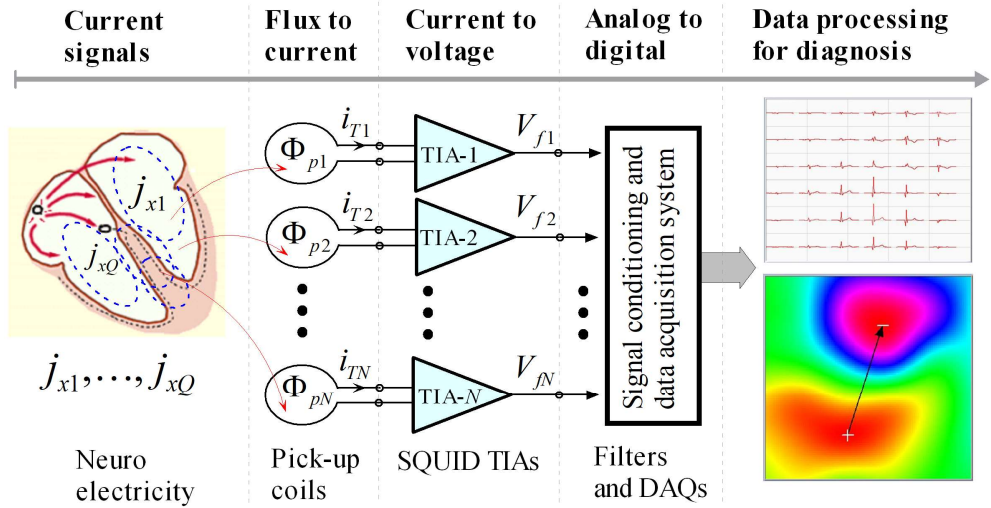

Fig. 46. A multichannel SQUID system for magnetocardiogram.

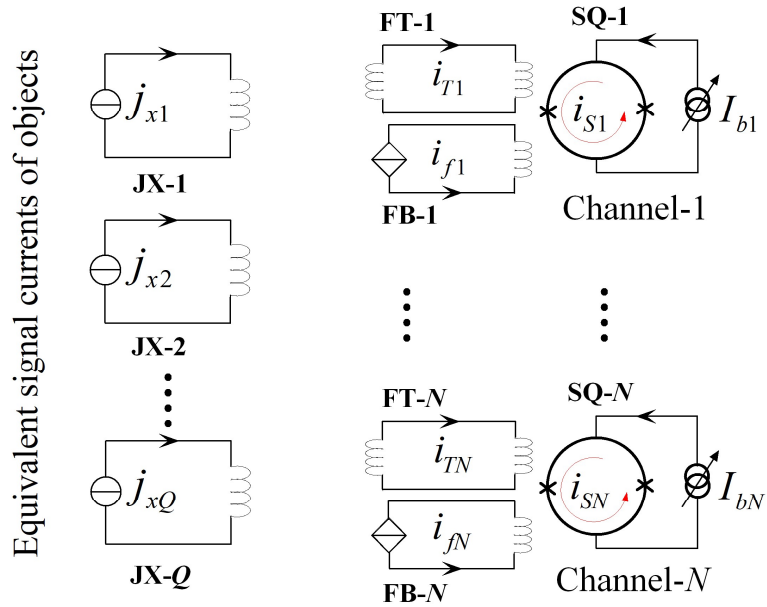

Fig. 47. General equivalent circuit of multi-channel SQUID sensors.

## *12.2 Equivalent circuit of multichannel SQUID system*

The equivalent circuit of the multichannel SQUID system is shown in Fig. 47. It is composed of five groups of mutually-coupled current loops.
1) The first group of current loops, JX-1, …, JX-$Q$, are assumed to generate the target magnetic fields coupled by the pick-up coils. They contain the currents circulating inside the object under test, namely $j_{x1}$, …, $j_{xQ}$.
2) The second group of current loops, FT-1, …, FT-$N$, are the flux transformers, their superconducting loop currents, $i_{T1}$, …, $i_{TN}$, are linearly induced by the fluxes coupled by the loops.

3) The third group of current loops, FB-1, …, FB-$N$, are the flux-feedback loops driven by the feedback currents, $i_{f1}$, …, $i_{fN}$.
4) The fourth group of current loops, SQ-1, …, SQ-$N$, are the dc-SQUID loops which inside contain superconducting currents, $i_{S1}$, …, $i_{SN}$.
5) The fifth group of current loops are the current-biasing loops driven by the external current sources, $I_{b1}$, …, $I_{bN}$.

(a)
$$\mathbf{M}_{xS} = \begin{matrix} & j_{x1} & \cdots & j_{xQ} \\ \text{SQ}-1 \\ \vdots \\ \text{SQ}-N \end{matrix} \begin{bmatrix} M_{xS11} & \cdots & M_{xS1Q} \\ \vdots & \ddots & \vdots \\ M_{xSN1} & \cdots & M_{xSNQ} \end{bmatrix}$$

$M_{xSij}$: mutual-inductance on SQ-$i$ by $j_{xj}$

(b)
$$\mathbf{M}_{fS} = \begin{matrix} & i_{f1} & \cdots & i_{fN} \\ \text{SQ}-1 \\ \vdots \\ \text{SQ}-N \end{matrix} \begin{bmatrix} M_{fS11} & \cdots & M_{fS1N} \\ \vdots & \ddots & \vdots \\ M_{fSN1} & \cdots & M_{fSNN} \end{bmatrix}$$

$M_{fSij}$: mutual-inductance on SQ-$i$ by $i_{fj}$

(c)
$$\mathbf{M}_{bS} = \begin{matrix} & I_{b1} & \cdots & I_{bN} \\ \text{SQ}-1 \\ \vdots \\ \text{SQ}-N \end{matrix} \begin{bmatrix} M_{bS11} & \cdots & M_{bS1N} \\ \vdots & \ddots & \vdots \\ M_{bSN1} & \cdots & M_{bSNN} \end{bmatrix}$$

$M_{bSij}$: mutual-inductance on SQ-$i$ by $I_{bj}$

(d)
$$\mathbf{M}_{TS} = \begin{matrix} & i_{T1} & \cdots & i_{TN} \\ \text{SQ}-1 \\ \vdots \\ \text{SQ}-N \end{matrix} \begin{bmatrix} M_{TS11} & \cdots & M_{TS1N} \\ \vdots & \ddots & \vdots \\ M_{TSN1} & \cdots & M_{TSNN} \end{bmatrix}$$

$M_{TSij}$: mutual-inductance on SQ-$i$ by $i_{Tj}$

Fig. 48. Mutual inductances applied to SQUIDs. (a) Mutual inductances from signal sources; (b) mutual inductances from feedback coils; (c) mutual inductances from bias currents; (d) mutual inductances from flux transformers.

(a)
$$\mathbf{M}_{xT} = \begin{matrix} & j_{x1} & \cdots & j_{xQ} \\ \text{FT}-1 \\ \vdots \\ \text{FT}-N \end{matrix} \begin{bmatrix} M_{xT11} & \cdots & M_{xT1Q} \\ \vdots & \ddots & \vdots \\ M_{xTN1} & \cdots & M_{xTNQ} \end{bmatrix}$$

$M_{xTij}$: mutual-inductance on FT-$i$ by $j_{xj}$

(b)
$$\mathbf{M}_{fT} = \begin{matrix} & i_{f1} & \cdots & i_{fN} \\ \text{FT}-1 \\ \vdots \\ \text{FT}-N \end{matrix} \begin{bmatrix} M_{fT11} & \cdots & M_{fT1N} \\ \vdots & \ddots & \vdots \\ M_{fTN1} & \cdots & M_{fTNN} \end{bmatrix}$$

$M_{fTij}$: mutual-inductance on FT-$i$ by $i_{fj}$

(c)
$$\mathbf{M}_{bT} = \begin{matrix} & I_{b1} & \cdots & I_{bN} \\ \text{FT}-1 \\ \vdots \\ \text{FT}-N \end{matrix} \begin{bmatrix} M_{bT11} & \cdots & M_{bT1N} \\ \vdots & \ddots & \vdots \\ M_{bTN1} & \cdots & M_{bTNN} \end{bmatrix}$$

$M_{bTij}$: mutual-inductance on FT-$i$ by $I_{bj}$

(d)
$$\mathbf{M}_{ST} = \begin{matrix} & i_{S1} & \cdots & i_{SN} \\ \text{FT}-1 \\ \vdots \\ \text{FT}-N \end{matrix} \begin{bmatrix} M_{TS11} & \cdots & M_{TS1N} \\ \vdots & \ddots & \vdots \\ M_{TSN1} & \cdots & M_{TSNN} \end{bmatrix} = \mathbf{M}_{TS}$$

$M_{TSij}$: mutual-inductance on FT-$i$ by $i_{Sj}$

(e)
$$\mathbf{L}_{T} = \begin{matrix} & i_{T1} & \cdots & i_{TN} \\ \text{FT}-1 \\ \vdots \\ \text{FT}-N \end{matrix} \begin{bmatrix} L_{T11} & \cdots & L_{T1N} \\ \vdots & \ddots & \vdots \\ L_{TN1} & \cdots & L_{TNN} \end{bmatrix}; L_{Tij} = \begin{cases} L_{pi} + L_{ii}, i = j: \text{self-inductance of FT-}i \\ -M_{Tij}, i \neq j: \text{mutual-inductance between FT-}i \text{ and FT-}j \end{cases}$$

Fig. 49. Mutual inductances applied to flux transformers. (a) Mutual inductances from signal sources; (b) mutual inductances from feedback coils; (c) mutual inductances from bias currents; (d) mutual inductances from SQUID loops; (e) mutual inductances between flux transformers.

Those loops are coupled through mutual inductances. Four mutual inductance matrices for the SQUID loops are defined in Fig. 48. Five mutual inductances for the flux transformers are defined in Fig. 49.

## 12.3 Transfer function of multichannel SQUID system

Three current vectors for the target currents, feedback currents, and bias currents, are defined in Fig. 50(a); the vectors of superconducting currents and trapped fluxes in flux transformers are defined in Fig. 50(b). The circulating currents and the applied fluxes inside SQUID loops are defined in Fig. 50(c), where the $i_S$ in a given dc-SQUID is decided by the $I_b$ and $\Phi_a$. The $\Phi_a$ achieves a fixed value as defined in (22) when the FLL is locked, and an unknown value when the FLL is unlocked.

**(a) Vector of current sources**

$$\mathbf{j}_x = \begin{bmatrix} j_{x1} & \cdots & j_{xQ} \end{bmatrix}^T$$

$$\mathbf{i}_f = \begin{bmatrix} i_{f1} & \cdots & i_{fN} \end{bmatrix}^T$$

$$\mathbf{I}_b = \begin{bmatrix} I_{b1} & \cdots & i_{bN} \end{bmatrix}^T$$

**(b) Vector of flux transformers**

$$\mathbf{i}_T = \begin{bmatrix} i_{T1} & \cdots & i_{TN} \end{bmatrix}^T$$

$$\mathbf{n}_T = \begin{bmatrix} n_{T1} & \cdots & n_{TN} \end{bmatrix}^T$$

**(c) Vector of SQUID loops**

$$\mathbf{i}_S = \begin{bmatrix} i_{S1} & \cdots & i_{SN} \end{bmatrix}^T; i_{Si} = \frac{1}{T}\int_0^T i_{ciri} \cdot dt = f_{ciri}(I_{bi},\Phi_{ai}); i \in [1,N]$$

$$\mathbf{\Phi}_a = \begin{bmatrix} \Phi_{a1} & \cdots & \Phi_{aN} \end{bmatrix}^T; \Phi_{ai} = \begin{cases} \Phi_{wi} + n_{wi}\Phi_0 - \Phi_{ni} & : \text{FLL is locked} \\ \text{unknown} & : \text{FLL is opened} \end{cases}$$

Fig. 50. (a) Current vectors of current sources. (b) current and flux vectors of flux transformers. (c) current and flux vectors of SQUIDs.

According to the FQL, the flux-current relations for flux transformers are written in matrix as

$$\mathbf{M}_{xT}\mathbf{j}_x + \mathbf{M}_{fT}\mathbf{i}_f + \mathbf{L}_T\mathbf{i}_T - \mathbf{M}_{bT}\mathbf{I}_b - \mathbf{M}_{TS}\mathbf{i}_S = \mathbf{n}_T\Phi_0 \qquad (45)$$

Meanwhile, the total fluxes applied to SQUID loops is expressed in matrix as

$$\mathbf{M}_{xS}\mathbf{j}_x + \mathbf{M}_{fS}\mathbf{i}_f - \mathbf{M}_{TS}\mathbf{i}_T - \mathbf{M}_{bS}\mathbf{I}_b = \mathbf{\Phi}_a \qquad (46)$$

Therefore, the transfer function of the mutual-channel SQUID system is derived as

$$\begin{aligned}\mathbf{i}_f = & \left(\mathbf{M}_{fS} + \mathbf{M}_{TS}\mathbf{L}_T^{-1}\mathbf{M}_{fT}\right)^{-1}\left(\mathbf{\Phi}_a + \mathbf{M}_{TS}\mathbf{L}_T^{-1}\mathbf{n}_T\Phi_0\right) \\ & + \left(\mathbf{M}_{fS} + \mathbf{M}_{TS}\mathbf{L}_T^{-1}\mathbf{M}_{fT}\right)^{-1}\mathbf{M}_{TS}\mathbf{L}_T^{-1}\mathbf{M}_{TS}\mathbf{i}_S \\ & + \left(\mathbf{M}_{fS} + \mathbf{M}_{TS}\mathbf{L}_T^{-1}\mathbf{M}_{fT}\right)^{-1}\left(\mathbf{M}_{bS} + \mathbf{M}_{TS}\mathbf{L}_T^{-1}\mathbf{M}_{bT}\right)\mathbf{I}_b \\ & - \left(\mathbf{M}_{fS} + \mathbf{M}_{TS}\mathbf{L}_T^{-1}\mathbf{M}_{fT}\right)^{-1}\left(\mathbf{M}_{xS} + \mathbf{M}_{TS}\mathbf{L}_T^{-1}\mathbf{M}_{xT}\right)\mathbf{j}_x\end{aligned} \qquad (47)$$

For SQUID sensors without flux transformers, the transfer function is simplified as

$$\mathbf{i}_f = \mathbf{M}_{fS}^{-1}\left(\mathbf{\Phi}_a + \mathbf{M}_{bS}\mathbf{I}_b - \mathbf{M}_{xS}\mathbf{j}_x\right) \qquad (48)$$

Therefore, the feedback currents are the linear projection of the target currents together with other loop currents. The crosstalk factors between SQUID sensors are induced by multiple mutual-inductance matrices. Since the calibration of those crosstalk factors through experiments is a heavy task [82], [83], the transfer functions in (47) and (48) can be used to calculate crosstalk factors with mutual inductances extracted according to the practical geometric dimensions of SQUID sensors.

## 13. Conclusion

We presented a comprehensive introduction to SQUID sensors, ranged from basic principles of superconductor elements to the general transfer function of multi-channel SQUID systems. SQUID sensors are the analog 'integrated' circuits of superconducting devices and semiconductor amplifiers working in different working temperatures. The understandings of SQUIDs and SQUID sensors are:

1) Superconductor and normal RLC circuits share the same circuit laws and analysis methods, except that the macroscopic phase or the integral of voltage will be used to define Josephson currents.
2) Flux-transformers are superconducting inductance loops wired by pure superconductor wires; SQUIDs are the superconducting RLC networks driven by Josephson currents. A flux-transformer is a linear flux-to-current convertor; a dc-SQUID is a MFET with flux-modulated current-voltage characteristics.
3) SQUID sensors are the linear flux- and current-based analog circuits, while MOSFET amplifiers are the linear charge- and voltage-based analog circuits.
4) A flux-follower implemented by the SQUID-based FF-OPAs is exactly the SQUID-based FLL.
5) A dc-SQUID can be read out by either CBVA or VBCA modes in a FLL. The VBCA mode will be better than the CBVA mode in selecting a working point.
6) Three near-SQUID feedback schemes are equivalent for improving the transfer coefficient of dc-SQUID.
7) A SQUID-based TIA is a flux-flower tightly coupled by an input coil. A SQUID sensor is a SQUID-based TIA connected to a pick-up coil.
8) A multi-channel SQUID sensor system, is a contactless current measurement system, where SQUID sensors read out the circuits circulating inside the target under test, through mutual inductances.

In summary, a dc-SQUID is a two-port RLC network driven by two Josephson currents; it is used as a MFET with flux-modulated current-voltage characteristics in FF-OPAs; a SQUID-based TIA is an FF-OPA coupled with an input coil; a SQUID sensor is A SQUID-based TIA connected with a pick-up coil. Multi-channel SQUID sensors are contactless current amplifiers. SQUID-based MFETs are dual to semiconductor FETs, in the design of analog circuits and systems.